\documentclass[11pt]{article}

\usepackage[margin=1in]{geometry}

\usepackage{amsmath,amsthm,amssymb}
\usepackage{ifthen}
\usepackage{thm-restate}
\usepackage{booktabs} 
\usepackage{graphicx}
\usepackage{float}
\usepackage{xfrac}
\usepackage{enumerate,multicol,multirow}
\usepackage{caption,subcaption}
\usepackage[hidelinks]{hyperref}
\usepackage{xurl}
\usepackage[round,authoryear]{natbib}
\usepackage{xcolor}
\usepackage[capitalize]{cleveref}
\crefname{observation}{Observation}{Observations}

\usepackage{tikz}
\usepackage{pgfplots} 
\pgfplotsset{compat=1.18}

\usepackage[ruled]{algorithm2e}

\SetAlFnt{\small}
\SetAlCapFnt{\small}
\SetAlCapNameFnt{\small}
\SetAlCapHSkip{0pt}
\IncMargin{-\parindent}

\usepackage{mdframed}

\DeclareMathOperator*{\argmax}{argmax}
\DeclareMathOperator{\poly}{poly}

\newtheorem{theorem}{Theorem}[section]
\newtheorem{lemma}[theorem]{Lemma}

\newtheorem{observation}[theorem]{Observation}

\newtheorem{proposition}[theorem]{Proposition}

\newcommand{\notshow}[1]{{}}

\newcommand{\Xcomment}[1]{{}}

\newcommand{\given}{\,\vert\,}

\newcommand{\prob}[2][]{\text{\bf Pr}\ifthenelse{\not\equal{}{#1}}{_{#1}}{}\!\left[{\def\givenn{\middle|}#2}\right]}
\newcommand{\expect}[2][]{\text{\bf E}\ifthenelse{\not\equal{}{#1}}{_{#1}}{}\!\left[{#2}\right]}

\newcommand{\tparen}{\big}
\newcommand{\tprob}[2][]{\text{\bf Pr}\ifthenelse{\not\equal{}{#1}}{_{#1}}{}\tparen[{\def\given{\tparen|}#2}\tparen]}
\newcommand{\texpect}[2][]{\text{\bf E}\ifthenelse{\not\equal{}{#1}}{_{#1}}{}\tparen[{\def\given{\tparen|}#2}\tparen]}

\newcommand{\sprob}[2][]{\text{\bf Pr}\ifthenelse{\not\equal{}{#1}}{_{#1}}{}[#2]}
\newcommand{\sexpect}[2][]{\text{\bf E}\ifthenelse{\not\equal{}{#1}}{_{#1}}{}[#2]}

\newcommand{\rbr}[1]{\left(#1\right)}

\newcommand{\reals}{\mathbb{R}}

\newcommand{\posterior}{\mu}

\newcommand{\sstate}{\omega}
\newcommand{\sstates}{\Omega}
\newcommand{\reward}{f}
\newcommand{\revenue}{g}
\newcommand{\agents}{[n]}

\newcommand{\contract}{\alpha}
\newcommand{\contractscheme}{\rho}

\newcommand{\cost}{c}
\newcommand{\OPT}{{\rm OPT}}
\newcommand{\Wel}{{\rm Wel}}

\newcommand{\lp}{\ell} % Notation for the indifference threshold
\newcommand{\prior}{\mathcal{D}}
\newcommand{\mean}{m}

\makeatletter
\AddToHook{cmd/appendix/before}{\def\cref@section@alias{appendix}\def\cref@subsection@alias{appendix}}

\begin{document}

\title{Transparency in Team Contracts}

\author{Paul D\"utting\thanks{Google Research, Z\"urich, Switzerland. Email: \texttt{duetting@google.com}} \and Yingkai Li\thanks{National University of Singapore, Singapore. Email: \texttt{yk.li@nus.edu.sg }} \and Inbal Talgam-Cohen\thanks{Tel Aviv University, Israel. Email: \texttt{italgam@tauex.tau.ac.il}} \and Jinzhao Wu\thanks{Yale University, New Haven, CT, USA. Email: \texttt{jinzhao.wu@yale.edu}}}

\date{}

\maketitle
\thispagestyle{empty}

\begin{abstract}
In contract design, the principal is often better informed about the project's chances of succeeding than the agents. To incentivize effort, the principal has two main tools: first, to design a contract, sharing part of the project's success among the agents; second, to design a (Bayesian) persuasion scheme, 
selectively sharing information with the agents. %
In this work we are interested in the optimal combination of both these tools. 

We tackle the joint problem through the prism of simplicity vs.~optimality, and 
establish a stark dichotomy: The simple disclosure scheme of full transparency 
achieves a tight $4/3$-approximation to the optimal combination of information disclosure and team contract, when the team's production function is submodular. In contrast, full transparency can be arbitrarily far from optimal when production is supermodular. 

The techniques that we develop to bound the gap for submodular functions, also yield a new deterministic polynomial-time approximation algorithm for pure multi-agent contract design, with a significantly improved approximation ratio of $2.311$, %for the %known-state 
and thus a $3.081$-approximation algorithm for the joint information and contracting problem.
\end{abstract}

\newpage
\setcounter{tocdepth}{2} 
\tableofcontents
\thispagestyle{empty}

\newpage

\pagestyle{plain}
\setcounter{page}{1}

%% Section: Introduction
\section{Introduction}
\label{sec:intro}

Algorithmic \emph{contract} design is an important frontier at the intersection of economics and computation \citep{duetting2024survey}. 
A fundamental problem in algorithmic contract design involves a principal who delegates a project to a team of agents, and seeks to design a (near-)optimal payment scheme to incentivize their costly and unobservable efforts~\citep{dutting2023multi}. 
In contracting problems, it is often the case that the principal is more informed than the agents. 
For example, the principal may hold information about the difficulty of the task, the state of the market, 
a competing lab's technology, etc. This information affects how the agents' efforts translate into success or failure of the project.
In such situations, in addition to designing the contract, 
the principal may gain from strategically revealing or withholding information. 
This brings us into the realm of algorithmic \emph{information} design --- another vibrant frontier~\citep{Dughmi2017Algorithmic,DughmiX21}. 

The combination of contract and information design introduces an intricate design problem of jointly optimizing the information disclosure and contracting schemes ---
see \citep{castiglioni2025hiring} for a recent algorithmic treatment in a single-agent environment. In general, joint optimization of two or more incentive schemes %among auction, contract and information design 
is typically complex, exhibiting all the known intricacies of multi-parameter auction design~\citep{Daskalakis2015MultiItem} (see~\citep{DaskalakisPT16} for the combination of auction and information design, and \citep{CastiglioniCLXZ25} for auction and contract design). Beyond presenting new technical challenges, the joint design of incentive schemes brings theory closer to real-world applications, where different mechanisms interact rather than operate in isolation.
The \emph{simple vs.~optimal} paradigm is particularly well-suited for bridging the gap between theory and practice, shedding light on why certain simple designs are ubiquitous despite their sub-optimality (see~\citep{BabaioffILW20} for simple auctions, and~\citep{dutting2019simple} for simple contracts). 

Among all possible complex information disclosure schemes, \emph{full transparency}, which fully reveals the principal's knowledge to the team of agents, stands out as simple and practical. 
In fact, disclosure of relevant business information is mandated in some settings and encouraged by common business wisdom~\citep{EU_Directive_2002_14_EC,BetrVG_English_2024,stack2013great,ohno2019toyota}. 
However, the economics literature has established that full transparency is strictly suboptimal in a wide range of settings~\citep{jehiel2015transparency,kwak2022optimal}, and so a theoretical explanation for the widespread adoption of full transparency remains lacking. Moreover, due to the complex nature of the problem, exact characterizations of optimal joint disclosure and contract schemes have remained elusive in many settings.
It thus remains unclear how much is lost due to full transparency, and whether (approximately-)optimal joint schemes can be efficiently computed for multi-agent settings. 

In this paper we answer these questions: We show that when team members are substitutes, a simple full-transparency scheme achieves a tight approximation of $\frac43$ relative to the optimal, potentially intricate joint design. We further provide an efficient algorithm for computing a full-revelation scheme coupled with contracts, which achieves a $3.081$-approximation to the optimum (a significant improvement over the previous best-known approximation for contracts alone).

\subsection{Our Results}

We start with the model of~\cite{dutting2023multi} in which the success of the project relies on the set of agents exerting effort on it, and add an unknown state-of-the-world $\omega\in\Omega$ drawn from a known prior. The principal jointly designs (i)~a signaling scheme that publicly informs all agents with certain information regarding the unknown state; and (ii)~a contract scheme with signal-dependent payments to the agents contingent on the project's success. The agents receive a signal and a contract, and choose whether or not to exert effort.

\paragraph{The Substitutes Case.}

\leavevmode 
We show that with substitutable (i.e., submodular) agents, full transparency provides a $\frac43$-approximation to the optimal joint information disclosure and contracting scheme. We thus find that while full transparency can be strictly suboptimal, for a broad class of multi-agent contract settings, it is \emph{approximately} optimal. 

\medskip

\noindent \textbf{Main Theorem 1} (\Cref{thm:submodular})\textbf{.}
For multi-agent settings with submodular success probability functions,
full transparency is a $\frac{4}{3}$-approximation to $\OPT$, and this is tight.

\medskip

The approximation ratio of $\frac43$ for submodular success probability functions holds regardless of the number of agents. It is approached by instances with \emph{additive} success probability functions, in which a different agent has a high contribution to success in each state. For the single-agent case, Appendix~\ref{sec:single-agent} improves this approximation ratio to the tight bound of $\frac98$.

Our approach for bounding the gap for submodular functions, 
also improves the computation of contracts. We give a deterministic value-oracle algorithm with approximation ratio $2.311$ for a known state. Combining this algorithm with full revelation gives a $3.081$-approximation to the optimal joint information and contracting scheme (\Cref{thm:computation}). The running time is polynomial in the number of agents, and the contract can be computed separately after each state is revealed. 
The algorithm and its proof appear in \Cref{app:algorithms}.

\paragraph{The Complements Case.}

For complementary agents, we find that coordination incentives reverse the previous conclusion: Pooling states together when disclosing state information can be valuable in ensuring and sustaining joint high-effort profiles. On the flip side, full transparency removes pooling across states and can inflate incentive costs. 
\medskip

\noindent \textbf{Main Theorem 2} (\Cref{thm:supermodular})\textbf{.} For multi-agent settings with complementary success probability functions,
the worst-case loss from full transparency compared to $\OPT$ is unbounded, even with only two agents.
\medskip

Specifically, we can construct environments where fully revealing information provides zero payoff for the principal, while providing no information leads to a strictly positive payoff. See Section~\ref{sec:complements} for our construction. 
Section~\ref{sec:conclude} concludes with open questions; we believe there is further potential for interesting research on private and multi-channel information structures for team contracts.

\subsection{Intuition for the Separation}

Before discussing the technical details, we provide some intuition regarding \emph{why} there is a separation between submodular and supermodular environments.

In general, withholding information from the agents can improve incentives. Conditional on inducing effort in every state, providing no information can allow the principal to implement effort at a strictly lower expected payment than under full transparency.

To illustrate this advantage in multi-agent environments, consider first the case of fixing the set of agents who are required to exert effort. Under full information, it may be necessary for the principal to transfer essentially all the surplus in every state in order to induce those agents to work. Under no information, by contrast, the principal may still be able to induce the same effort profile while retaining a positive fraction of the generated surplus. In this sense, the gap between full and no information can be arbitrarily large: full revelation can drive the principal's residual payoff to zero. This {phenomenon} does not depend on whether the success probability function is submodular or supermodular. 

Where submodularity and supermodularity begin to matter is when the principal is allowed to choose which agents to incentivize in each state. With submodular success probabilities, the principal can start from the team she would incentivize under no information and, in each realized state, dismiss agents whose contribution in that state is low and hence costly to incentivize. Submodularity ensures that dismissing an agent can only make the remaining agents cheaper to incentivize, and by carefully choosing whom to dismiss, the principal retains at least a $\frac34$-fraction of her no-information utility in expectation. Economically speaking, when agents are substitutes, dismissing or not incentivizing a small number of agents does not dramatically harm output.

With supermodular success probability functions, by contrast, effort is complementary. Inducing effort from only a subset of agents can cause a drastic collapse in success probabilities. In fact, when the state is fully revealed, it may be that for every state and every nontrivial subset of agents, inducing effort requires giving away the entire surplus. This complementarity is the core reason we obtain sharply different approximation guarantees in the submodular versus supermodular cases. 

\subsection{Our Techniques}

A key ingredient of all our approximation results is a result we refer to as the ``All-or-Nothing'' Lemma (\cref{lem:approximation_extension}). This result shows that the worst-case approximation of full transparency compared to the optimal scheme is attained by its worst case approximation to no transparency. 
This significantly simplifies our analysis, as contract design under zero transparency is analogous to contract design without states, where the principal's utility guarantees are well-understood.

The All-or-Nothing Lemma greatly simplifies the analysis of the multi-agent setting. Building on this lemma, all we need to show for our positive result for submodular rewards (\Cref{thm:submodular}) is that designing contracts separately for each state of the world is sufficient to guarantee a good approximation to the no information optimal contract in expectation.

We establish the tight bound of $\frac43$ via a \emph{Potential Lemma} (\Cref{lem:pruning}). Starting from the optimal team under no information, in each realized state we process the agents one at a time, in decreasing order of the ratio between their realized and expected marginal contributions, and decide whether to keep or delete each agent. Agents whose realized marginal contribution is at least their expected one are always kept, since they are cheaper to incentivize than under no information. The delicate decisions concern agents with a low realized marginal contribution: keeping them is costly, whereas deleting them reduces the success probability (although, by submodularity, it makes the remaining agents cheaper to incentivize). We guide these decisions by a potential function, which combines the principal's residual share, the current success probability, and a correction term that vanishes in expectation over the states. We then show that a guarantee never decreases during the pruning process and is achieved by the final team. This guarantee consists of the savings from kept agents with high realized marginal contributions plus the positive part of the potential. The key calculation shows that, for an agent whose realized marginal contribution is a $t$-fraction of its expected one, a $t:(1-t)$ average of keeping and deleting the agent loses at most a $t(1-t)\le\frac14$ fraction of its expected marginal contribution. This is exactly where the factor $\frac43$ comes from. A matching lower bound is attained by instances with additive success probability functions.

For the computational result, we strengthen the pruning analysis to control both the reward lost and the sum of the square roots of the retained agents' costs. This permits repeated greedy refilling of a pruned team to a fixed cost budget. Every unsuccessful round increases the reward by a definite amount, which bounds the number of rounds. A quadratic potential makes the pruning decisions deterministic. We then enumerate the agents with large singleton incentive shares and use cost scaling to extend the guarantee to arbitrary instances, retaining any positive baseline success probability.

Finally, to show the negative result for supermodular rewards (\Cref{thm:supermodular}), we devise an explicit setting with two agents and two states of the world where full revelation achieves zero utility, while no disclosure achieves a strictly positive principal's utility.

\subsection{Related Work}

Multi-agent contract settings were introduced by~\cite{holmstrom1982moral} and considered from a combinatorial perspective by \citep{babaioff2012combinatorial}. We adopt the model introduced by \citep{dutting2023multi}. Recent advances in the algorithmic study of multi-agent contract settings include
\citep{CastiglioniM023,cacciamani2024multi,deo2024supermodular,DFG24,dutting2025multi,alon2025multi,doron2026algorithm,bollini2026matching,FeldmanSS26,anagnostides2026compensation}. For other combinatorial contract settings see \cite{DuttingRT21,DuttingEFK21,feldman2025combinatorialcontractdesignrecent}.
For recent work on algorithmic information design see \cite{Xu20,GradwohlHHS21}.

The role of transparency and contracting abilities in organizational settings has been discussed extensively in the economics literature \citep[e.g.,][]{holmstrom1979moral,rayo2010optimal,jehiel2015transparency}. 
Most prominently, \citep{jehiel2015transparency} shows that if the principal can contract on the unknown state, full transparency is generically suboptimal when the state space is sufficiently rich. 
However, if the principal can only contract based on the communicated signals instead of the unknown state, \citep{kwak2022optimal} shows that whether full transparency is optimal depends on the order of the increase in the success probability by exerting effort, and the order of the rate of increase across states. 
This dependency on the ability to contract on the unknown state is also echoed by the recent work of \citep{curello2024complete}. 
Our paper focuses on the latter model, 
and shows that full transparency achieves a tight $\frac43$-approximation in multi-agent environments with a substitutable (i.e., submodular) production function. 

The most relevant work that focus on computational aspects is 
\citep{castiglioni2025hiring}, which considers the computation of the joint design of information structures and contracts in a single-agent environment. 
They show that when the principal can contract on the state, the problem of computing the optimal contract admits an additive FPTAS through linear programming, and this is essentially the best possible, as the supremum principal's utility may be unattainable. However, when the principal lacks that contracting power, computing the optimal solution is APX-hard. Another closely related work is \cite{BabichenkoTXZ24}, which considers the joint design of information and contracts in the single-agent setting and characterizes implementable actions and utility profiles. A main difference of our work is that we focus on simple, robust designs and consider multi-agent settings.

The multi-agent environments we consider in our paper entail both submodularity \citep{dutting2023multi} and supermodularity \citep{deo2024supermodular}, where computing the exact optimal solution is hard.
For submodular rewards in the known-state setting, \citep{dutting2023multi} established constant-factor approximation algorithms, and the concurrent work of \citep{FeldmanSS26} improved the value-oracle approximation factor to $6.128$. Our computational result further improves this factor to $2.311$ and also allows for a positive reward for the empty team.
Our paper also connects to the economics literature on team production with complements \citep[e.g.,][]{itoh1991incentives,kandel1992peer,halac2021rank} and substitutes \citep[e.g.,][]{babaioff2012combinatorial,bag2019input}.
Their main focus is on the impact of complementarity or substitutability on the form of optimal contracts, 
and on economic implications such as free-riding effects. 
In contrast, our paper focuses on the role of complementarity or substitutability in the approximation guarantees of full transparency. Our results for complement agents relate to the insight that more precise public information can reduce welfare when agents also have private information \citep{morris2002social}. 

In relation to \citep{Aharoni0T25,FeldmanTPS25}, our positive results have the following interesting consequence: full revelation is approximately optimal not just for maximizing the principal's utility, but also for maximizing the \emph{second-best} welfare in a multi-agent contract setting. The second-best welfare is the optimal welfare attainable by a \emph{feasible} contract~$\contract$ that satisfies $\sum_{i\in\agents}{\contract_i}\le 1$ (the agents' shares in the project cannot exceed $1$). In environments with unknown states, approximate optimality does not imply that full transparency increases second-best welfare. Our results establish approximate optimality for submodular rewards through the constant-factor connection between the optimal principal's utility and the second-best welfare from~\citep{Aharoni0T25,FeldmanTPS25}.

%% Section: Model
\section{Model}
\label{sec:model}

\paragraph{Multi-Agent Contract Settings with States.}

Our starting point is the binary-action binary-outcome multi-agent contracts setting of \cite{dutting2023multi}, but we augment this model with an uncertain state of the world $\omega \in \Omega$, which determines how effort translates into success. We assume that $|\Omega|$ is finite.

A principal (she, her) seeks to delegate the execution of a project to the agents. 
We use $j \in \{0,1\}$ to denote the outcome, where $0$ stands for the project's failure and $1$ for its success. The principal's reward in the case of success is normalized to $1$, and is $0$ otherwise. 

Each agent $i \in [n]$ (he, his) privately chooses whether to exert effort or not. The cost of effort for agent $i$ is $c_i\in \reals_+$, and otherwise, his cost is normalized to $0$. We denote the set of agents who exert effort by $S\subseteq[n]$. Importantly, 
the agents' efforts are hidden; the principal does not observe~$S$, and only learns whether or not the project is eventually successful.

A \emph{success probability} set function $f_\omega: 2^{[n]} \rightarrow [0,1]$, which depends on the state of the world $\omega \in \Omega$, maps each set of agents $S$ who exert effort to a probability $f_\omega(S) \in [0,1]$ with which the project succeeds. Since we normalize the principal's reward for success to $1$, the expected reward of the principal is also $f_\omega(S)$. We therefore sometimes refer to $f_\omega$ as the \emph{reward} function, \emph{production} function, or \emph{technology} (it maps effort to reward). 
We assume that $f_\omega$ is monotone for each $\omega \in \Omega$, meaning that $f_\omega(S) \leq f_\omega(T)$ whenever $S \subseteq T$. In addition, we often impose additional structure on $f_\omega$, such as submodularity or supermodularity.%
\footnote{We generally allow for $f_\omega(\emptyset) > 0$.}
Formally, given a set function $f$ over $[n]$, an agent $i \in [n]$, and a set $S \subseteq [n]$, let $f(i \mid S) := f(S \cup \{i\}) - f(S)$ denote the \emph{marginal contribution} of agent $i$ to $S$. We consider the following classes of set functions:
\begin{itemize}
\item Submodular: A set function $f$ is \emph{submodular} if for any two sets $S \subseteq S'$ and any $i \in [n]$ it holds that $f(i \mid S) \geq f(i \mid S')$.
\item Supermodular: A set function $f$ is \emph{supermodular} if for any two sets $S \subseteq S'$ and any $i \in [n]\setminus S'$ it holds that $f(i \mid S) \leq f(i \mid S')$.
\end{itemize}

\paragraph{Information Design.}

We explore situations with information asymmetry between a principal and agents regarding the state of the world. We focus on the case in which the principal is better informed than the agents.
Adopting the Bayesian persuasion framework of \cite{rayo2010optimal,KamenicaG11}, we assume that the principal observes the state $\omega$, while the agents only have distributional information in the form of a prior distribution $\mathcal{D}\in \Delta(\Omega)$ from which the state of the world is drawn.

The principal strategically reveals information to the agents by committing to a \emph{signaling scheme} $\pi: \Omega \rightarrow \Delta(\Sigma)$, where $\Sigma$ is a compact space of signals. 
The signaling scheme is known to the agents. Upon observing the realized state $\omega$, the principal draws a signal $\sigma$ from the distribution $\pi(\omega)$. The realized signal $\sigma$ is observed by all agents, i.e., the signaling scheme is \emph{public}. The agents use Bayes' rule to update their belief about the state of the world to a \emph{posterior} distribution $\mu(\sigma)$, common to all agents. 
Since $\pi$ induces a system of (probabilistic) beliefs, we also sometimes use the term \emph{information structure} to refer to the signaling scheme. 

\paragraph{Contract Design.} 

To incentivize the agents to exert effort, in addition to the signaling scheme, the principal designs a 
\emph{contract scheme} $\contractscheme: \Sigma \rightarrow [0,1]^n$,
which maps a signal $\sigma$ to a \emph{contract} $\contract = \contractscheme(\sigma) \in [0,1]^n$. 
For every $i\in[n]$, $\contract_i$ is the fraction (a.k.a.~\emph{share}) of the reward that agent $i$ receives if the project is successful.%
\footnote{Equivalently, $\contract_i$ is the fraction of the principal's expected reward transferred to agent $i$. A contract that specifies the agent's fraction is known as \emph{linear}; such contracts are without loss of generality in binary-outcome settings.}
We denote by $\contractscheme_i : \Sigma \to [0,1]$ the mapping of the signal to~$\contract_i$.
Like the signaling scheme, the contract scheme $\contractscheme$ is public information. So, given a signal $\sigma$, each agent $i$ can infer the shares $\contractscheme_{i'}(\sigma)$ of all agents $i' \in [n]$. 
A subtle but important point is the dependence of $\contractscheme$ on the signal rather than directly on the state of the world. This means the agents have certainty about the contract when choosing whether or not to exert effort, and that the state of the world need not be contractible. 

\paragraph{Timeline.} 
Our contract setting is summarized by a tuple $(\mathcal{D},\{f_\omega\},c)$: a prior $\mathcal{D}$ over the states; state-dependent reward functions $\{f_\omega\}_{\omega\in\Omega}$, each of which is a set function over $[n]$; and costs $c=(c_1,\dots,c_n)$. The setting is commonly known. For simplicity, we sometimes omit components of the setting from the notation where clear from context. The timing of the game is as follows:

\begin{enumerate}
    \item The principal publicly commits to a pair $(\pi,\contractscheme)$ of signaling scheme $\pi$ 
    and contract scheme $\contractscheme$. 
    \item The state of the world $\omega$ is drawn from $\mathcal{D}$ and observed by the principal.
    \item The principal draws a signal $\sigma\sim \pi(\omega)$ according to the signaling scheme and broadcasts it to the agents. 
    \item The agents update their belief about the state of the world from prior $\mathcal{D}$ to posterior $\mu(\sigma)$. Moreover, given that they observed the signal $\sigma$, the agents also know that they will be compensated via contract $\contractscheme(\sigma)$.
    \item \label{item:game} Each agent $i \in [n]$ decides whether to exert effort or not. 
    Let $S$ be the set of agents who exert effort; $S$ is unknown to the principal.
    Each agent $i \in S$ incurs a cost $c_i$. 
    \item The 
    project succeeds with probability $f_\omega(S)$. In case of success, the principal receives a reward of $1$ and pays $\contractscheme_i(\sigma)$ to each agent $i \in [n]$. Otherwise, the principal receives no reward, and the agents are not compensated. 
\end{enumerate}

Note that if the state of the world is known (e.g., if $\mathcal D$ is a point-mass distribution), we obtain a ``stateless'' multi-agent contract setting $(f,c)$ as in \cite{dutting2023multi}.

\paragraph{Utilities.}

In expectation over the project's success, the utility of agent $i$ under contract $\contract \in [0,1]^n$ depends on the state of the world $\omega$ and the set of agents $S \subseteq [n]$ who exert effort, and is given by
\[
    U_i(S, \contract, \omega) := \contract_i \cdot f_\omega(S) - \mathbf{1}\{i \in S\} \cdot c_i.
\]
Similarly, the principal's utility is given by 
\[
    U_P(S, \contract, \omega) := \rbr{1-\sum_{i \in [n]} \contract_i} \cdot f_\omega(S).
\]
Given a distribution $\mu\in\Delta(\Omega)$ over the states of the world, the (expected) utilities are
\[
U_i(S, \contract, \posterior) := \expect[\omega \sim \mu]{U_i(S, \contract, \omega)};~~~
U_P(S, \contract, \posterior) := \expect[\omega \sim \mu]{U_P(S, \contract, \omega)}.
\]

\paragraph{Equilibrium Concept.} 

With public signals and expected-utility maximizing agents, the game played in Stage~\ref{item:game} is equivalent to a \emph{stateless} multi-agent contract game. For exposition, we formulate the model and analysis using pure Nash equilibria (\textsf{NE}). All our bounds on the price of transparency extend with the same approximation factors when both full transparency and the optimal public information scheme are evaluated under %principal-optimal 
mixed Nash equilibria; see \Cref{app:BeyondPureNash} for the reduction and proofs.

Formally, fix a posterior $\mu$ and a contract $\contract$. (For example, $\mu = \mu(\sigma)$ and $\contract = \contractscheme(\sigma)$.) 
A set $S\subseteq[n]$ of agents exerting effort induces utility $U_i(S, \contract, \posterior)$ for every agent $i$. Then $S$ forms an equilibrium (denoted $S\in\textsf{NE}(\contract,\posterior)$) if for all $i \in [n]$:
\[
U_i(S, \contract, \posterior) \ge \max\big\{U_i(S\setminus\{i\}, \contract, \posterior),~U_i(S\cup\{i\}, \contract, \posterior)\big\}.
\]
Note that we can write $U_i(S, \contract, \posterior)=\contract_i \cdot \expect[\omega \sim \mu]{f_{\omega}(S)} - \mathbf{1}\{i \in S\} \cdot c_i$.
So $S$ is just a standard pure Nash equilibrium of contract $\contract$ in the stateless multi-agent contract game $(f_\mu,c)$, with reward function $f_\mu$ defined by $f_{\mu}(T) := \expect[\omega \sim \mu]{f_{\omega}(T)}$ for each set $T \subseteq [n]$, and with costs $c=(c_1, \ldots, c_n)$. It is well-known that in such stateless multi-agent contract games, any contract $\contract$ admits a pure Nash equilibrium.%
\footnote{This follows from a potential game argument \citep{deo2024supermodular,dutting2025multi}. To show existence, fix agents with $\contract_i=0$ to shirk and maximize $f_\mu(S)-\sum_{i\in S}\frac{c_i}{\contract_i}$ over $S \subseteq \{i \in [n]: \alpha_i > 0\}$.}

\paragraph{The Joint Design Problem.} The principal's goal (which we adopt) is the joint design of a signaling 
and contract scheme
$(\pi,\contractscheme)$, which maximizes the principal's expected utility under the principal-optimal equilibrium. 

It is again instructive to adopt a posterior view. Given any posterior belief $\posterior$, the principal's utility under the optimal pure Nash equilibrium given contract $\contract$ is
\[
U_P(\contract,\posterior) := \max_{S \in \text{\textsf{NE}}(\contract,\posterior)} U_P(S,\contract,\posterior). 
\]
So, for a fixed prior $\mathcal{D}$, the principal's a priori expected utility from a joint signaling-contract scheme $(\pi,\contractscheme)$ is 
\[
U_P(\contractscheme,\pi):= \expect[\omega\sim\prior,\sigma\sim \pi(\omega)]{U_P(\contractscheme(\sigma),\posterior(\sigma))}.
\]
We can now express the principal's goal as finding $(\pi^*,\contractscheme^*) \in \argmax_{\pi,\contractscheme} U_P(\contractscheme,\pi)$,\footnote{Since $\Omega$ is finite, \citet{KamenicaG11} imply that the supremum is attained by a joint scheme with finitely many signals.}
and define
\[
\OPT(\prior) := U_P(\contractscheme^*,\pi^*)
\] as the corresponding optimal utility. 

An alternative view of the principal's goal is that, given the prior $\prior$, she seeks to induce the best possible distribution over posterior beliefs, which we denote by $\pi\given \prior$. Each posterior belief $\mu$ corresponds to a principal's utility, namely the utility $\max_\alpha U_P(\alpha, \mu)$ that the principal can extract from the stateless contract game given by $(f_\mu,c)$. The principal seeks to maximize this utility in expectation when $\mu$ is drawn from $\pi\given \prior$.

%% Section: Structural Insights
\section{Structural Insights}
\label{sec:structure}

\paragraph{Full Transparency/Opaqueness.}
One extremal information structure is the fully-revealing (or fully transparent) signaling scheme $\pi^F$ with $\Sigma = \Omega$, which, for each $\omega \in \Omega$, assigns all probability to~$\omega$. 
Under this scheme, the posterior belief of the agents has no residual uncertainty. Let $\delta_{\omega}$ denote the degenerate probability distribution with all mass on $\omega$.  Using this, the principal's utilities 
given contract scheme $\contractscheme:\Sigma\to[0,1]^n$ and, respectively, given the optimal contract scheme are
\[
U_P^F(\contractscheme,\prior) := \expect[\omega \sim \mathcal{D}]{U_P(\contractscheme(\omega), \delta_{\omega})};
~~~U_P^F(\prior) := \max_{\contractscheme:\Sigma\to[0,1]^n} U_P^F(\contractscheme,\prior).
\] 

At the other end of the spectrum are the non-revealing (fully opaque) information structures, where $\Sigma = \{\sigma^*\}$ and the principal sends an uninformative signal $\sigma^*$ given all states. Under such a scheme, which we denote by $\pi^N$, the posterior $\mu(\sigma^*)$ is identical to the prior $\mathcal {D}$. It is then without loss for the principal to use a degenerate contract scheme, which specifies a single contract given signal $\sigma^*$.
The principal's utilities given contract $\contract\in [0,1]^n$ and, respectively, given the optimal contract are 
\[
U_P^N(\contract,\prior) := U_P(\contract,\prior);
~~~U_P^N(\prior) := \max_{\contract \in [0,1]^n} U_P^N(\alpha,\mathcal{D}). 
\]

\paragraph{The All-or-Nothing Lemma.}

We now state a simple but important lemma showing it is sufficient to compare full revelation only to the no information (fully opaque) signaling scheme.
This reduction greatly simplifies our analyses in subsequent sections. 

More precisely, in subsequent sections we aim to bound $\sup_{\mathcal{D}} \OPT(\mathcal{D})/U_P^F(\mathcal{D})$. It is clear that $\sup_{\mathcal{D}} \OPT(\mathcal{D})/U_P^F(\mathcal{D}) \geq \sup_{\mathcal{D}} U_P^N(\mathcal{D})/U_P^F(\mathcal{D})$. The following lemma shows that 
\[
\sup_{\mathcal{D}} \OPT(\mathcal{D})/U_P^F(\mathcal{D}) \leq \sup_{\mathcal{D}} U_P^N(\mathcal{D})/U_P^F(\mathcal{D}).
\]
So $\sup_{\mathcal{D}} \OPT(\mathcal{D})/U_P^F(\mathcal{D}) = \sup_{\mathcal{D}} U_P^N(\mathcal{D})/U_P^F(\mathcal{D})$.

\begin{lemma}[The All-or-Nothing Lemma]
\label{lem:approximation_extension}
Let $\beta\geq 1$ be an approximation factor. If full revelation is a $\beta$-approximation to no information for every contract setting $(\mathcal{D},\{f_\omega\},c)$, 
then full revelation is a $\beta$-approximation to the \emph{optimal} information structure. 
\end{lemma}

\begin{proof}
Fix a contract setting $(\mathcal{D},\{f_\omega\},c)$.
Let $\pi^*$ be the optimal signaling scheme and let $\pi^*\given\prior$ be the distribution it induces over posteriors. 
Since full revelation is a $\beta$-approximation to no information, this holds in particular for every contract setting $(\mu,\{f_\omega\},c)$ where $\mu\in\Delta(\Omega)$ is a posterior, and so 
\begin{equation}
    U^F_P(\mu)\ge \frac{1}{\beta}U^N_P(\mu).\label{eq:F-N}
\end{equation}

Let $\OPT(f_\omega,c)$ denote the principal's utility from choosing, at state $\omega$, the optimal contract for the stateless contract setting $(f_\omega,c)$. Then with full revelation, $U^F_P(\mu)=\expect[\omega\sim\mu]{\OPT(f_\omega,c)}$ for every distribution $\mu\in\Delta(\Omega)$ over states.  
Since by the law of iterated expectations, the expected posterior induced by a signaling scheme is equal to the prior, 
drawing $\omega$ from $\prior$ is equivalent to drawing a posterior $\mu$ from $\pi^*\given \prior$ and then drawing $\omega$ from $\mu$. So 
\begin{equation}
    U^F_P(\prior)=\expect[\omega\sim\prior]{\OPT(f_\omega,c)}=\expect[\mu\sim\pi^*\given\prior,\omega\sim\mu]{\OPT(f_\omega,c)}=\expect[\mu\sim \pi^*\given\prior]{U^F_P(\mu)}.\label{eq:full-prior-posterior}
\end{equation}

Putting everything together, 
\[
U^F_P(\prior)
\;=\;
\expect[\mu\sim \pi^*\given \prior]{U^F_P(\mu)}
\;\ge\;
\frac{1}{\beta}\,\expect[\mu\sim \pi^*\given \prior]{U^N_P(\mu)}
\;=\;
\frac{1}{\beta}\,\OPT(\prior),%\qedhere
\]
where the first equality is by \eqref{eq:full-prior-posterior} and the inequality is by \eqref{eq:F-N}. The last equality is since $U_P^N(\mu)=\OPT(f_\mu,c)$ (with no revelation the principal plays in the stateless game $(f_\mu,c)$), and since by definition of $\pi^*$ it holds that
$\expect[\mu\sim \pi^*\given \prior]{\OPT(f_\mu,c)}=\OPT(\prior)$.
\end{proof}

%% Section: Transparency with Substitution
\section{Transparency with Substitution}
\label{sec:multi-agents}

In this section, we show that the approximate optimality of full transparency extends to multi-agent settings if the project exhibits a substitution effect, i.e., if the reward functions are submodular. We pin down the worst-case loss exactly: full transparency is a $\frac43$-approximation to the optimal information structure, regardless of the number of agents, and this is tight.

\begin{theorem}
\label{thm:submodular}
For every multi-agent contract setting with states, where the reward function $\reward_{\sstate}$ is submodular for every state $\sstate \in \sstates$, it holds that full revelation is a $\frac{4}{3}$-approximation to the optimal information structure. Moreover, the factor $\frac{4}{3}$ is tight, even if the reward function $\reward_\sstate$ is additive for every state $\sstate\in\sstates$.
\end{theorem}

The instances that approach the ratio of $\frac43$ in \Cref{thm:submodular} have a growing number of agents.

In the remainder of the section we prove \Cref{thm:submodular}. The proof relies on the fact that, for submodular rewards, the marginal contribution of an agent can only increase when other agents are removed from the team.

\paragraph{Minimum-Payment Contracts.}
Consider a stateless multi-agent contract setting $(\reward,\cost)$, and let $\OPT(\reward,\cost)$ denote the principal's optimal utility in it. Suppose the principal wants exactly the agents in a team $S\subseteq\agents$ to exert effort.
Recall that $\reward(i\mid A):=\reward(A\cup\{i\})-\reward(A)$ denotes the marginal contribution of agent $i$ to a set $A$. An agent $i\in S$ weakly prefers exerting effort, with utility $\contract_i\reward(S)-\cost_i$, to free-riding on the rest of the team, with utility $\contract_i\reward(S\setminus\{i\})$, if and only if $\contract_i\,\reward(i\mid S\setminus\{i\})\ge\cost_i$. The lowest such share is therefore $\contract_i=\cost_i/\reward(i\mid S\setminus\{i\})$, where by convention $\contract_i=0$ if $\cost_i=0$.
Every agent $i\notin S$ receives $\contract_i=0$ and thus weakly prefers not to exert effort.
We call this contract the \emph{minimum-payment contract} for $S$. Whenever all its shares are at most $1$, it is feasible and $S$ is one of its pure Nash equilibria.
We denote its total share by $\kappa(S)$ and the resulting (possibly negative) principal's utility by $\revenue(S)$:
\[
\kappa(S)=\sum_{i\in S}\frac{\cost_i}{\reward(i\mid S\setminus\{i\})}
\qquad\text{and}\qquad
\revenue(S)=\big(1-\kappa(S)\big)\,\reward(S).
\]%
Every contract under which $S$ is an equilibrium pays each agent $i\in S$ at least its minimum-payment share. Hence $\OPT(\reward,\cost)$ is the maximum of $\revenue(S)$ over all teams $S$ whose minimum-payment contract is feasible.

\paragraph{Overview of the Proof of \texorpdfstring{\Cref{thm:submodular}}{Theorem 5.1}.}
By the All-or-Nothing Lemma (\Cref{lem:approximation_extension}), it suffices to compare full transparency with no disclosure. Under no disclosure, the principal faces the stateless setting $(\reward_\prior,\cost)$, where $\reward_\prior(S)=\expect[\sstate\sim\prior]{\reward_\sstate(S)}$, and optimally incentivizes some team $S^\star$, paying each agent $i\in S^\star$ with $\cost_i>0$ the share $\kappa_i=\cost_i/d_i$, where $d_i=\reward_\prior(i\mid S^\star\setminus\{i\})$ is agent $i$'s \emph{expected} marginal contribution (agents with zero cost receive zero shares). Under full transparency, the principal could keep incentivizing $S^\star$ in every state. However, in a realized state $\sstate$, the marginal contribution $\reward_\sstate(i\mid S^\star\setminus\{i\})$ of an agent may be far below its expected value $d_i$, in which case incentivizing the agent requires a share much larger than $\kappa_i$. Dismissing such an agent saves its share but reduces the success probability; at the same time, by submodularity, it increases the marginal contributions---and thus decreases the required shares---of the agents who remain.

Our main tool, the \emph{Potential Lemma} (\Cref{lem:pruning}), resolves this trade-off state by state. In every realized state $\sstate$, it lower-bounds the principal's optimal utility by
$(1-\kappa(S^\star))\big(\reward_\sstate(S^\star)-\frac14\sum_{i}d_i\big)$, plus a correction term that is linear in the realized marginal contributions $\reward_\sstate(i\mid S^\star\setminus\{i\})$ and vanishes in expectation over the states. By submodularity, $\sum_i d_i\le\reward_\prior(S^\star)$, so in expectation full transparency recovers at least a $\frac34$-fraction of the no-disclosure utility. We prove the Potential Lemma via an explicit, deterministic procedure that processes the agents one at a time and decides whether to keep or dismiss each of them, guided by a potential function; the savings from kept agents plus the positive part of this potential form a guarantee that never decreases along the procedure. The constant $\frac14$ equals $\max_{t\in[0,1]}t(1-t)$: it is the worst-case loss when deciding whether to keep or dismiss an agent whose realized marginal contribution is a $t$-fraction of its expected marginal contribution.

\subsection{The Potential Lemma}
\label{sub:pruning}

The Potential Lemma is a statement about a stateless setting $(\reward,\cost)$. In the proof of \Cref{thm:submodular}, we apply it separately in each realized state $\sstate$, with $\reward=\reward_\sstate$ (after folding zero-cost agents into the baseline), using the shares and the expected marginal contributions under no disclosure as a benchmark.

\begin{lemma}[Potential Lemma]
\label{lem:pruning}
Consider a stateless multi-agent contract setting $(\reward,\cost)$ with a monotone submodular reward function $\reward$, and a (possibly empty) team $S\subseteq\agents$ in which every agent has a strictly positive cost. Let $(\kappa_i)_{i\in S}$ be arbitrary strictly positive \emph{benchmark shares} with $\sum_{i\in S}\kappa_i\le 1$, and let $d_i:=\cost_i/\kappa_i$ be the corresponding \emph{benchmark marginals}. For each $i\in S$, let $p_i:=\reward(i\mid S\setminus\{i\})$ denote agent $i$'s marginal contribution to $S$. Then
\begin{equation}\label{eq:pruning}
\OPT(\reward,\cost)\;\ge\;\Big(1-\sum_{i\in S}\kappa_i\Big)\Big(\reward(S)-\frac14\sum_{i\in S}d_i\Big)+\sum_{i\in S}\kappa_i\,(p_i-d_i).
\end{equation}
\end{lemma}

The benchmark shares should be thought of as the shares paid under no disclosure, and the benchmark marginals as the corresponding expected marginal contributions. In particular, the right-hand side is affine in $\reward$, and when $d_i$ is the expectation of $p_i$ over the states, the correction term vanishes in expectation. Formally, if we apply the lemma in each state $\sstate$ with $\reward=\reward_\sstate$ and the same benchmark shares, and $d_i=\expect[\sstate\sim\prior]{\reward_\sstate(i\mid S\setminus\{i\})}$, then
\[
\expect[\sstate\sim\prior]{\sum_{i\in S}\kappa_i\big(\reward_\sstate(i\mid S\setminus\{i\})-d_i\big)}=0.
\]%
Note that the benchmark shares are parameters of the lemma, and need not coincide with the minimum-payment shares $\cost_i/p_i$ for $S$ under $\reward$; in our application, they are the minimum-payment shares under no disclosure, so that $\sum_{i}\kappa_i=\kappa(S^\star)$ with $\kappa(\cdot)$ defined with respect to $\reward_\prior$.

\begin{proof}
We first describe the idea of the proof. Start from the team $S$ and offer each agent $i$ its benchmark share $\kappa_i$. Since $\kappa_i d_i=\cost_i$, this share makes agent $i$ willing to exert effort exactly when its marginal contribution to the current team $T$, $m_i=\reward(i\mid T\setminus\{i\})$, is at least $d_i$. Agents with $m_i\ge d_i$ can thus be kept at share at most $\kappa_i$. An agent with $m_i=t\,d_i$ for some $t<1$ needs the larger share $\cost_i/m_i=\kappa_i/t$. Suppose we keep such an agent with probability $t$, paying $\kappa_i/t$, and delete it with probability $1-t$, paying nothing but losing its marginal contribution $m_i=t\,d_i$. Then the expected payment to agent $i$ is $t\cdot\kappa_i/t=\kappa_i$, exactly its benchmark share, while the expected loss is $(1-t)\,t\,d_i\le\frac14 d_i$. The proof below derandomizes these coin flips: it processes the agents one at a time and, guided by a potential function, makes a deterministic choice that is at least as good as the random one.

Write $N:=|S|$ and $B_0:=1-\sum_{i\in S}\kappa_i\ge 0$. Since $\cost_i>0$ and $\kappa_i>0$, we have $d_i>0$ for every $i\in S$. For a team $T$ and an agent $i\in T$, we refer to $\reward(i\mid T\setminus\{i\})$ as $i$'s \emph{current marginal} in $T$; it is nonnegative since $\reward$ is monotone. Submodularity of $\reward$ implies that removing other agents from a team can only increase the current marginal of an agent who remains:
\begin{equation}\label{eq:prune-DR}
\reward(i\mid T\setminus\{i\})\;\ge\;\reward(i\mid T'\setminus\{i\}) \qquad \text{whenever } i\in T\subseteq T'.
\end{equation}
In particular, $\reward(i\mid T\setminus\{i\})\ge p_i$ for every team $T\subseteq S$ that contains $i$.

\paragraph{The Pruning Procedure.}
We analyze a deterministic procedure (Algorithm~\ref{alg:pruning}) that processes the agents in $S$ one at a time and decides, for each agent, whether to keep or delete it. After $k\in\{0,\ldots,N\}$ agents have been processed, the agents in $S$ are partitioned into four sets: the agents $H_k$ kept as \emph{high} agents, the agents $L_k$ kept as \emph{low} agents, the deleted agents $D_k$, and the remaining (not yet processed) agents $R_k$. The current team is $T_k:=H_k\cup L_k\cup R_k=S\setminus D_k$. For each agent $i$, let $m_i$ denote its current marginal at the time it is processed. Since the procedure only deletes agents, \eqref{eq:prune-DR} implies that at every later stage, the current marginal of a kept agent $i$ is at least $m_i$.

The procedure is formally described in Algorithm~\ref{alg:pruning}. It starts by reserving the benchmark share $\kappa_i$ for every agent $i\in S$, which leaves the principal the \emph{residual share} $B=1-\sum_{i\in S}\kappa_i\ge0$. It then processes the agents from the best-performing one down: in each iteration, it selects a remaining agent $i$ with the highest ratio $t=m_i/d_i$ between its current marginal $m_i$ and its benchmark marginal $d_i$.

\emph{High agents.} If $t\ge1$, the reserved share suffices, since $\kappa_i m_i\ge\kappa_i d_i=\cost_i$, and agent $i$ is kept as a high agent.

\emph{Low agents.} If $t<1$, agent $i$ needs the larger share $\cost_i/m_i=\kappa_i/t$, its minimum share in the current team, which also suffices for $i$ in every later team by~\eqref{eq:prune-DR}. The procedure can either keep $i$ as a low agent, paying the extra share $\cost_i/m_i-\kappa_i$ out of the residual share, or delete $i$, which returns $\kappa_i$ to the residual share but lowers the reward by $m_i$. The procedure keeps $i$ if this results in a weakly higher potential (defined below) than deleting $i$, provided that keeping $i$ is affordable, i.e., $t>0$ and $B+\kappa_i-\cost_i/m_i\ge0$, so that the total share stays at most $1$. Otherwise, it deletes $i$.

\emph{The potential.} The decisions are guided by
\begin{equation}\label{eq:potential}
\Phi(B,T,R):=B\,\Big(\reward(T)-\frac14\sum_{j\in R}d_j\Big)+\sum_{j\in R}\kappa_j\,(p_j-d_j),
\end{equation}
a pessimistic estimate of the principal's final utility, given the residual share $B$, the current team $T$, and the set $R$ of remaining agents. Initially, $\Phi$ equals the right-hand side of~\eqref{eq:pruning}; at the end, $R=\emptyset$ and $\Phi=B\,\reward(T)$ bounds the principal's utility from below. The proof shows that $\Phi$, together with the savings from high agents, never decreases.

\begin{mdframed}[nobreak=true]
\begin{algorithm}[H]
\caption{\label{alg:pruning} Pruning of team \(S\) given benchmark shares \((\kappa_i)_{i\in S}\).}
\(T \gets S\), \(R \gets S\), \(H,L,D \gets \emptyset\), \(B \gets 1-\sum_{i\in S}\kappa_i\)\;
\While{\(R \neq \emptyset\)}{
Select \(i\in R\) maximizing \(\reward(i\mid T\setminus\{i\})/d_i\), breaking ties by a fixed rule\;
\(m_i \gets \reward(i\mid T\setminus\{i\})\), \(t\gets m_i/d_i\), \(R\gets R\setminus\{i\}\)\;
\uIf{\(t\ge 1\)}{
\(H\gets H\cup\{i\}\) \tcp*[r]{keep \(i\) as a high agent}
}
\uElseIf{\(t>0\), \(B+\kappa_i-\cost_i/m_i\ge 0\), and \(\Phi(B+\kappa_i-\cost_i/m_i,\,T,\,R)\ge\Phi(B+\kappa_i,\,T\setminus\{i\},\,R)\)}{
\(L\gets L\cup\{i\}\), \(B\gets B+\kappa_i-\cost_i/m_i\) \tcp*[r]{keep \(i\) as a low agent}
}
\Else{
\(D\gets D\cup\{i\}\), \(T\gets T\setminus\{i\}\), \(B\gets B+\kappa_i\) \tcp*[r]{delete \(i\)}
}
}
\Return team \(T\) with its minimum-payment contract \(\contract\)\;
\end{algorithm}
\end{mdframed}

It remains to show that the contract returned by Algorithm~\ref{alg:pruning} achieves the bound in~\eqref{eq:pruning}. For $k\in\{0,\ldots,N\}$, let $B_k$ denote the value of $B$ after $k$ agents have been processed. By the updates in Algorithm~\ref{alg:pruning},
\begin{equation}\label{eq:prune-B}
B_k=1-\sum_{h\in H_k}\kappa_h-\sum_{\ell\in L_k}\frac{\cost_\ell}{m_\ell}-\sum_{j\in R_k}\kappa_j.
\end{equation}
That is, $B_k$ is the share left to the principal if every high agent and every remaining agent is paid its benchmark share, and every low agent is paid its minimum share at the time it was kept. We further define
\begin{equation}\label{eq:prune-QG}
Q_k:=\sum_{h\in H_k}\kappa_h\,(m_h-d_h),\qquad \Phi_k:=\Phi(B_k,T_k,R_k),\qquad G_k:=Q_k+\max\{\Phi_k,0\}.
\end{equation}
Since every high agent $h$ satisfies $m_h\ge d_h$, the term $Q_k\ge 0$ accumulates the savings from high agents, which are cheaper to incentivize than their benchmark shares suggest. The quantity $G_k$ is the guarantee that the procedure maintains. Initially, $T_0=R_0=S$ and $H_0=L_0=D_0=\emptyset$, so $Q_0=0$ and $\Phi_0$ equals the right-hand side of~\eqref{eq:pruning}. Hence $G_0\ge \Phi_0$, and the lemma follows from the following two claims, which we prove in two steps below:
\begin{enumerate}[(i)]
\item $B_k\ge 0$ for every $k\le N$, and $G_{k+1}\ge G_k$ for every $k<N$;
\item the contract returned by Algorithm~\ref{alg:pruning} is feasible, and gives the principal a utility of at least $G_N$.
\end{enumerate}
Indeed, (i) and (ii) imply that $\OPT(\reward,\cost)\ge G_N\ge G_0\ge\Phi_0$.

\paragraph{Step 1: The Guarantee Never Decreases.}
First, $B_k\ge0$ for every $k$: initially $B_0\ge0$; keeping a high agent leaves $B$ unchanged, deleting an agent increases $B$, and a low agent is kept only if the updated value of $B$ is nonnegative.

Now fix $k<N$, and let $i$ be the agent processed in iteration $k+1$, with current marginal $m_i=\reward(i\mid T_k\setminus\{i\})$ and ratio $t=m_i/d_i\ge0$. By~\eqref{eq:prune-DR},
\begin{equation}\label{eq:prune-p}
p_i\;\le\; m_i \;=\; td_i.
\end{equation}
We separate agent $i$ from the other remaining agents by defining
\[
X:=\reward(T_k)-\frac14\sum_{j\in R_k\setminus\{i\}}d_j \qquad\text{and}\qquad Y:=\sum_{j\in R_k\setminus\{i\}}\kappa_j\,(p_j-d_j),
\]
which do not depend on the decision regarding agent $i$. By definition,
\begin{equation}\label{eq:prune-parent}
\Phi_k=B_k\,\Big(X-\frac{d_i}{4}\Big)+\kappa_i\,(p_i-d_i)+Y.
\end{equation}
If agent $i$ is kept as a low agent, the team is unchanged, and the residual share becomes $B_k+\kappa_i-\cost_i/m_i=B_k+\kappa_i-\kappa_i/t$, where we used $\cost_i=\kappa_i d_i=\kappa_i m_i/t$. If agent $i$ is deleted, the residual share becomes $B_k+\kappa_i$, and the reward decreases by $\reward(T_k)-\reward(T_k\setminus\{i\})=m_i=td_i$. Since $i$ is no longer a remaining agent in either case, let $\Phi^{\rm keep}$ denote the potential after keeping $i$ as a low agent (if $t>0$), and $\Phi^{\rm del}$ the potential after deleting $i$. By the definitions of $X$ and $Y$,
\begin{equation}\label{eq:prune-children}
\Phi^{\rm keep}=\Big(B_k+\kappa_i-\frac{\kappa_i}{t}\Big)X+Y \qquad\text{and}\qquad \Phi^{\rm del}=(B_k+\kappa_i)\,(X-td_i)+Y.
\end{equation}
We distinguish three cases.

\smallskip
\noindent\emph{Case 1: $t\ge 1$ (agent $i$ is high).} Agent $i$ is kept as a high agent. The team and the residual share are unchanged, $T_{k+1}=T_k$ and $B_{k+1}=B_k$ (the term $\kappa_i$ moves from the last sum in~\eqref{eq:prune-B} to the first). Since agent $i$ is no longer a remaining agent,
\[
\Phi_{k+1}-\Phi_k=\frac{B_kd_i}{4}-\kappa_i\,(p_i-d_i)\qquad\text{and}\qquad Q_{k+1}-Q_k=\kappa_i\,(m_i-d_i)\;\ge\; 0.
\]
Adding these two equations, and using $B_k\ge0$ and~\eqref{eq:prune-p}, we obtain
\[
(Q_{k+1}+\Phi_{k+1})-(Q_k+\Phi_k)=\frac{B_kd_i}{4}+\kappa_i\,(m_i-p_i)\;\ge\;0.
\]
Since $G_k=\max\{Q_k,\,Q_k+\Phi_k\}$ and both arguments of the maximum weakly increase, $G_{k+1}\ge G_k$.

\smallskip
\noindent\emph{Case 2: $0<t<1$, and keeping $i$ is affordable, i.e., $B_k+\kappa_i-\kappa_i/t\ge 0$.} In this case, Algorithm~\ref{alg:pruning} chooses the option with the higher potential, so $\Phi_{k+1}=\max\{\Phi^{\rm keep},\Phi^{\rm del}\}$, while $Q_{k+1}=Q_k$. To show that $\max\{\Phi^{\rm keep},\Phi^{\rm del}\}\ge\Phi_k$, we consider the convex combination with weight $t$ on keeping and $1-t$ on deleting. These weights are chosen so that the residual shares average to the current one:
$t\,(B_k+\kappa_i-\kappa_i/t)+(1-t)\,(B_k+\kappa_i)=B_k$. Hence, by~\eqref{eq:prune-children},
\[
t\,\Phi^{\rm keep}+(1-t)\,\Phi^{\rm del}=B_kX-(B_k+\kappa_i)\,d_i\,t(1-t)+Y.
\]
Subtracting~\eqref{eq:prune-parent}, we obtain
\begin{align}
t\,\Phi^{\rm keep}+(1-t)\,\Phi^{\rm del}-\Phi_k
&{}=B_kd_i\,\Big(\frac14-t(1-t)\Big)+\kappa_i\,\big(d_i-p_i-d_it(1-t)\big)\nonumber\\
&{}\ge B_kd_i\,\Big(t-\frac12\Big)^2+\kappa_i d_i\,(1-t)^2\;\ge\;0,\label{eq:prune-avg}
\end{align}
where the first inequality uses $\frac14-t(1-t)=(t-\frac12)^2$ and $d_i-p_i-d_it(1-t)\ge d_i-td_i-d_it(1-t)=d_i(1-t)^2$, by~\eqref{eq:prune-p}. Since a maximum is at least any convex combination, $\Phi_{k+1}\ge\Phi_k$, and thus $G_{k+1}\ge G_k$.

\smallskip
\noindent\emph{Case 3: $t<1$, and keeping $i$ is not affordable, i.e., $t=0$ or $B_k+\kappa_i-\kappa_i/t<0$.} In this case, agent $i$ is deleted, so $\Phi_{k+1}=\Phi^{\rm del}$ and $Q_{k+1}=Q_k$. If $\Phi_k\le 0$, then $G_{k+1}\ge Q_{k+1}=Q_k=G_k$. Suppose therefore that $\Phi_k>0$. In both subcases, the fact that keeping $i$ is not affordable can be written as
\begin{equation}\label{eq:prune-unaffordable}
\kappa_i\,(1-t)\;>\;B_kt.
\end{equation}
Indeed, for $t=0$, \eqref{eq:prune-unaffordable} holds since $\kappa_i>0$; for $t>0$, it follows from $B_k+\kappa_i-\kappa_i/t<0$ by multiplying by $t$.
Here we use the order in which the agents are processed: since agent $i$ maximizes the ratio between current marginal and benchmark marginal among all remaining agents, every $j\in R_k$ satisfies $\reward(j\mid T_k\setminus\{j\})\le td_j<d_j$, and hence $p_j<d_j$ by~\eqref{eq:prune-DR}. Therefore, $Y\le 0$. Moreover, $d_i-p_i\ge d_i(1-t)>0$ by~\eqref{eq:prune-p}. Together with $\Phi_k>0$ and~\eqref{eq:prune-parent}, this gives
\[
B_k\,\Big(X-\frac{d_i}{4}\Big)\;>\;\kappa_i\,(d_i-p_i)-Y\;\ge\;\kappa_i\,(d_i-p_i)\;>\;0.
\]
Since $B_k\ge0$, it follows that $B_k>0$, and
\begin{equation}\label{eq:prune-X}
X\;>\;\frac{d_i}{4}+\frac{\kappa_i}{B_k}\,(d_i-p_i).
\end{equation}
Using~\eqref{eq:prune-parent} and~\eqref{eq:prune-children}, and then~\eqref{eq:prune-X},
\begin{align*}
\Phi^{\rm del}-\Phi_k
&{}=\kappa_i X+\frac{B_kd_i}{4}+\kappa_i\,(d_i-p_i)-(B_k+\kappa_i)\,td_i\\
&{}>(B_k+\kappa_i)\,d_i\,\Big[\frac14-t+\frac{\kappa_i}{B_k}\Big(1-\frac{p_i}{d_i}\Big)\Big]\\
&{}\ge(B_k+\kappa_i)\,d_i\,\Big[\frac14-t+\frac{\kappa_i}{B_k}\,(1-t)\Big]\;>\;\frac{(B_k+\kappa_i)\,d_i}{4},
\end{align*}
where the second inequality uses~\eqref{eq:prune-p}, and the last inequality uses~\eqref{eq:prune-unaffordable}. Hence $\Phi_{k+1}=\Phi^{\rm del}>\Phi_k$, and $G_{k+1}\ge G_k$.

\paragraph{Step 2: The Final Contract Achieves the Guarantee.}
After all $N$ agents have been processed, no agent remains to be processed, i.e., $R_N=\emptyset$, and the final team $T_N=H_N\cup L_N$ consists of the kept high and low agents. By~\eqref{eq:potential},
\[
\Phi_N=\Phi(B_N,T_N,\emptyset)=B_N\Big(\reward(T_N)-\frac14\sum_{j\in\emptyset}d_j\Big)+\sum_{j\in\emptyset}\kappa_j\,(p_j-d_j)=B_N\,\reward(T_N)\ge0,
\]
where the inequality uses $B_N\ge0$ (Step~1) and $\reward\ge0$. Hence
\[
G_N=Q_N+B_N\,\reward(T_N).
\]
For each $i\in T_N$, let $\bar m_i:=\reward(i\mid T_N\setminus\{i\})$ denote its final marginal. As argued above, $\bar m_i\ge m_i$; moreover, $m_h\ge d_h>0$ for every high agent $h\in H_N$, and $m_\ell>0$ for every low agent $\ell\in L_N$, since a low agent is kept only if $t>0$. Hence the minimum-payment contract $\contract$ for $T_N$ returned by Algorithm~\ref{alg:pruning}, with $\contract_i=\cost_i/\bar m_i$ for $i\in T_N$ and $\contract_i=0$ for $i\notin T_N$, is well defined. By~\eqref{eq:prune-B} with $R_N=\emptyset$,
\begin{equation}\label{eq:prune-decomp}
1-\sum_{i\in T_N}\contract_i\;=\;B_N+\sum_{h\in H_N}\big(\kappa_h-\contract_h\big)+\sum_{\ell\in L_N}\Big(\frac{\cost_\ell}{m_\ell}-\contract_\ell\Big).
\end{equation}
Every summand on the right-hand side is nonnegative: for $h\in H_N$, we have $\contract_h=\cost_h/\bar m_h\le \cost_h/d_h=\kappa_h$, since $\bar m_h\ge m_h\ge d_h$; and for $\ell\in L_N$, we have $\contract_\ell=\cost_\ell/\bar m_\ell\le\cost_\ell/m_\ell$. Together with $B_N\ge 0$, this shows that $\sum_{i\in T_N}\contract_i\le 1$. Hence every share lies in $[0,1]$, so the contract $\contract$ is feasible, and $T_N$ is a pure Nash equilibrium of $\contract$, since every agent in $T_N$ is indifferent between exerting effort and not, and every agent outside $T_N$ has a zero share.

It remains to bound the principal's utility from $T_N$. For each $h\in H_N$, using $\contract_h=\kappa_h d_h/\bar m_h$ and $\reward(T_N)\ge \reward(T_N)-\reward(T_N\setminus\{h\})=\bar m_h$ (as $\reward$ is nonnegative),
\[
(\kappa_h-\contract_h)\,\reward(T_N)=\kappa_h\Big(1-\frac{d_h}{\bar m_h}\Big)\reward(T_N)\;\ge\;\kappa_h\Big(1-\frac{d_h}{\bar m_h}\Big)\bar m_h=\kappa_h\,(\bar m_h-d_h)\;\ge\;\kappa_h\,(m_h-d_h).
\]
Multiplying~\eqref{eq:prune-decomp} by $\reward(T_N)\ge0$, dropping the nonnegative terms of the low agents, and summing the last inequality over $h\in H_N$, we conclude that
\[
\OPT(\reward,\cost)\;\ge\;\Big(1-\sum_{i\in T_N}\contract_i\Big)\reward(T_N)\;\ge\;B_N\,\reward(T_N)+Q_N\;=\;G_N.
\]
Combined with Step~1, $\OPT(\reward,\cost)\ge G_N\ge G_0\ge\Phi_0$, which proves the lemma.
\end{proof}

The order in which the agents are processed matters only in Case~3: always processing a remaining agent with the largest ratio between its current and benchmark marginals (recomputed after every deletion) guarantees that whenever keeping a low agent is not affordable, all remaining agents are low as well.

\subsection{Proof of \texorpdfstring{\Cref{thm:submodular}}{Theorem 5.1}}
\label{sub:proof-submodular}

We prove the upper bound and the tightness of the approximation factor $\frac43$ separately.

\paragraph{Upper Bound.}
We show that full revelation is a $\frac43$-approximation to the optimal information structure.

\begin{proof}[Proof of the upper bound in \Cref{thm:submodular}]
Fix a multi-agent contract setting $(\prior,\{\reward_\sstate\},\cost)$ in which every $\reward_\sstate$ is submodular. By the All-or-Nothing Lemma (\Cref{lem:approximation_extension}), it suffices to show that $U^F_P(\prior)\ge\frac34\,U^N_P(\prior)$. Under no disclosure, the principal faces the stateless setting $(\reward_\prior,\cost)$, where $\reward_\prior(S)=\expect[\sstate\sim\prior]{\reward_\sstate(S)}$ is monotone and submodular, as a convex combination of monotone submodular functions. Let $S^\star$ be an optimal team for $(\reward_\prior,\cost)$, so that $U^N_P(\prior)=\OPT(\reward_\prior,\cost)=\revenue(S^\star)=(1-\kappa(S^\star))\,\reward_\prior(S^\star)$, where $\kappa$ and $\revenue$ are defined with respect to $\reward_\prior$. If $U^N_P(\prior)=0$, there is nothing to prove, so assume that $U^N_P(\prior)>0$, which implies $\kappa(S^\star)<1$.

Partition $S^\star$ into the agents $P:=\{i\in S^\star:\cost_i>0\}$ with positive costs and the agents $Z:=S^\star\setminus P$ with zero costs. For $i\in P$, let
\[
d_i:=\reward_\prior(i\mid S^\star\setminus\{i\})\qquad\text{and}\qquad \kappa_i:=\frac{\cost_i}{d_i}
\]
denote agent $i$'s expected marginal contribution and its share under no disclosure, respectively. Since $S^\star$ is incentivized by a contract with shares in $[0,1]$, we have $d_i\ge\cost_i>0$; and since zero-cost agents receive zero shares, $\sum_{i\in P}\kappa_i=\kappa(S^\star)<1$.

To deal with the zero-cost agents, we fold them into the baseline: for each state $\sstate$, define $\reward^Z_\sstate(A):=\reward_\sstate(A\cup Z)$ for $A\subseteq\agents$. The function $\reward^Z_\sstate$ is monotone and submodular, with $\reward^Z_\sstate(P)=\reward_\sstate(S^\star)$ and $\reward^Z_\sstate(i\mid P\setminus\{i\})=\reward_\sstate(i\mid S^\star\setminus\{i\})$ for every $i\in P$. Moreover,
\begin{equation}\label{eq:fold-zero-cost}
\OPT(\reward_\sstate,\cost)\;\ge\;\OPT(\reward^Z_\sstate,\cost).
\end{equation}
To see this, let $A$ be a pure Nash equilibrium of a contract $\contract$ in the setting $(\reward^Z_\sstate,\cost)$. Then $A\cup Z$ is a pure Nash equilibrium of $\contract$ in the setting $(\reward_\sstate,\cost)$: every agent $i\notin Z$ faces the same utilities from exerting effort and from not exerting effort in both settings, and for every agent $i\in Z$, exerting effort is a best response, since $\cost_i=0$ and $\reward_\sstate$ is monotone. Since $\reward_\sstate(A\cup Z)=\reward^Z_\sstate(A)$, the principal's utility is the same in both settings, which proves~\eqref{eq:fold-zero-cost}.

Now fix a state $\sstate$, and apply the Potential Lemma (\Cref{lem:pruning}) to the setting $(\reward^Z_\sstate,\cost)$ with team $P$ and benchmark shares $(\kappa_i)_{i\in P}$, whose benchmark marginals are $\cost_i/\kappa_i=d_i$. Together with~\eqref{eq:fold-zero-cost}, this gives
\begin{equation}\label{eq:statewise}
\OPT(\reward_\sstate,\cost)\;\ge\;\big(1-\kappa(S^\star)\big)\Big(\reward_\sstate(S^\star)-\frac14\sum_{i\in P}d_i\Big)+\sum_{i\in P}\kappa_i\,\big(\reward_\sstate(i\mid S^\star\setminus\{i\})-d_i\big).
\end{equation}
We take expectations over $\sstate\sim\prior$ on both sides of~\eqref{eq:statewise}. The last sum vanishes in expectation, since $\expect[\sstate\sim\prior]{\reward_\sstate(i\mid S^\star\setminus\{i\})}=\reward_\prior(i\mid S^\star\setminus\{i\})=d_i$ for every $i\in P$. Recalling that $U^F_P(\prior)=\expect[\sstate\sim\prior]{\OPT(\reward_\sstate,\cost)}$, we obtain
\[
U^F_P(\prior)\;\ge\;\big(1-\kappa(S^\star)\big)\Big(\reward_\prior(S^\star)-\frac14\sum_{i\in P}d_i\Big).
\]
Finally, order the agents in $S^\star$ arbitrarily. By submodularity of $\reward_\prior$, the marginal contribution of each agent $i\in S^\star$ to $S^\star\setminus\{i\}$ is at most its marginal contribution to the agents preceding it, and the latter marginal contributions sum to $\reward_\prior(S^\star)-\reward_\prior(\emptyset)$. Hence, $\sum_{i\in P}d_i\le\sum_{i\in S^\star}\reward_\prior(i\mid S^\star\setminus\{i\})\le\reward_\prior(S^\star)$. Therefore,
\[
U^F_P(\prior)\;\ge\;\frac34\,\big(1-\kappa(S^\star)\big)\,\reward_\prior(S^\star)\;=\;\frac34\,U^N_P(\prior).\qedhere
\]
\end{proof}

The proof of the upper bound is constructive: given the optimal team $S^\star$ under no disclosure, running Algorithm~\ref{alg:pruning} on $\reward^Z_\sstate$ in each realized state $\sstate$ (and paying zero shares to the agents in $Z$) yields a full-transparency contract scheme whose expected utility is at least $\frac34\,U^N_P(\prior)$. In each state, the procedure makes $|P|$ decisions and uses $O(|P|^2)$ evaluations of $\reward_\sstate$. This does not by itself provide an efficient algorithm, as it requires knowledge of $S^\star$.

\paragraph{Tightness.}
The following proposition shows that the factor $\frac43$ in \Cref{thm:submodular} cannot be improved, even for additive reward functions. Since additive functions are submodular, this completes the proof of \Cref{thm:submodular}.

\begin{proposition}
\label{prop:submodular-tight}
For every even $n\ge 2$, there exists a multi-agent contract setting with $n$ agents, $n$ states, and additive reward functions, in which
\[
\frac{\OPT(\prior)}{U^F_P(\prior)}\;\ge\;\frac{U^N_P(\prior)}{U^F_P(\prior)}\;=\;\frac{4(3n^3+n-1)}{3n^2(3n+2)}.
\]
In particular, the lower bound on $\OPT(\prior)/U^F_P(\prior)$ approaches $\frac43$ as $n\to\infty$, which matches the upper bound of \Cref{thm:submodular}.
\end{proposition}

\begin{proof}
Let $M:=2n^2+n+1$, and note that $(2n-1)M=4n^3+n-1$. The states $\sstate\in\sstates=\agents$ are equally likely, and every agent has the same cost $\cost_i=c:=\frac{n}{(2n-1)M}$. In state $\sstate$, agent $\sstate$ is the \emph{high} agent, and the reward function is
\[
\reward_\sstate(S)=\overline{x}\cdot\mathbf 1\{\sstate\in S\}+\underline{x}\cdot|S\setminus\{\sstate\}|,\qquad\text{where }\ \overline{x}:=\frac{n}{2n-1}\ \text{ and }\ \underline{x}:=\frac{1}{2n-1}.
\]
Each $\reward_\sstate$ is additive, with $\reward_\sstate(\emptyset)=0$ and $\reward_\sstate(\agents)=\overline{x}+(n-1)\,\underline{x}=1$. In an additive setting, the marginal contribution of an agent does not depend on the rest of the team. Moreover, the minimum-payment shares of every team sum to less than $1$, both under no disclosure and in every revealed state (by the minimum shares computed below, these sums are at most $n^2c=\frac{n^3}{(2n-1)M}<1$ and $\frac{1+n(n-1)}{M}<1$, respectively). Hence, $\OPT(\reward,\cost)=\max_S\revenue(S)$, where the maximum ranges over all teams $S$.

\smallskip
\noindent\emph{No disclosure.} Every agent is high in exactly one of the $n$ states, so $\reward_\prior(S)=|S|\cdot\frac{\overline{x}+(n-1)\,\underline{x}}{n}=\frac{|S|}{n}$. Thus every agent's minimum share is $nc<1$, and a team of $s$ agents yields $(1-s\cdot nc)\,\frac{s}{n}=\frac sn-s^2c$. This is a concave function of $s$, whose derivative at $s=n$ is $\frac1n-2nc>0$, since $(2n-1)M=4n^3+n-1>2n^3$. Hence, the optimal team under no disclosure consists of all agents, and
\[
U^N_P(\prior)=1-n^2c=\frac{3n^3+n-1}{(2n-1)M}.
\]

\smallskip
\noindent\emph{Full transparency.} Fix a revealed state. The high agent's minimum share is $c/\overline{x}=\frac1M$, and every other agent's minimum share is $c/\underline{x}=\frac nM$. Hence, a team consisting of the high agent and $s\in\{0,\ldots,n-1\}$ other agents, and a team consisting of $s\in\{0,\ldots,n-1\}$ agents other than the high agent, respectively yield
\[
\Big(1-\frac{1+ns}{M}\Big)\frac{n+s}{2n-1}=\frac{n\,(2n+1-s)(n+s)}{(2n-1)M}
\qquad\text{and}\qquad
\Big(1-\frac{ns}{M}\Big)\frac{s}{2n-1}=\frac{s\,(M-ns)}{(2n-1)M}.
\]
The first expression is a concave function of $s$, maximized at $s=\frac{n+1}{2}$; since $n$ is even, its maximum over integers is attained at $s=\frac n2$, with value $\frac{3n^2(3n+2)}{4(2n-1)M}$. The second expression is increasing in $s\le n-1$, since its unconstrained maximizer is $\frac{M}{2n}>n$; hence, it is at most $\frac{(n-1)(n+1)^2}{(2n-1)M}$, which is smaller than the former value, as
\[
\frac{3n^2(3n+2)}{4}-(n-1)(n+1)^2=\frac{5n^3+2n^2+4n+4}{4}>0.
\]
Therefore, in every state an optimal team consists of the high agent and $\frac n2$ other agents, and
\[
U^F_P(\prior)=\frac{3n^2(3n+2)}{4(2n-1)M}.
\]
Since no disclosure is a feasible information structure, $\OPT(\prior)\ge U^N_P(\prior)$. Dividing the expressions for $U^N_P(\prior)$ and $U^F_P(\prior)$ yields the claimed ratio, and $\frac43-\frac{4(3n^3+n-1)}{3n^2(3n+2)}=\frac{4(2n^2-n+1)}{3n^2(3n+2)}\to0$ as $n\to\infty$.
\end{proof}

\subsection{\texorpdfstring{Computational Implications}{Computational Implications}}
\label{sub:computation}

The proof of \Cref{thm:submodular} relies on a pruning technique that selects which agents to retain after the state is revealed. Applying this technique to contract computation also yields improved approximation algorithms. We first obtain a guarantee for computing a contract when the state is known, and then combine it with \Cref{thm:submodular} to approximate the joint design problem.

\begin{theorem}[Computing contracts and full-disclosure schemes]\label{thm:computation}
Let $\varpi\in(0,1)$ be the unique solution of $\varpi e^{\varpi}=1$, and set $\gamma_\star=1-\varpi$. For every fixed $\varepsilon>0$, the following guarantees hold.
\begin{enumerate}[(i)]
\item \textbf{Known state.} For every stateless multi-agent contract setting $(\reward,\cost)$ with a monotone submodular reward function, a deterministic algorithm computes a contract and an equilibrium team whose principal utility is at least
\[
\frac{\OPT(\reward,\cost)}{\gamma_\star^{-1}+\varepsilon},
\qquad
\gamma_\star^{-1}=2.3102333355\ldots.
\]
\item \textbf{Uncertain state.} If $\reward_{\sstate}$ is submodular for every $\sstate\in\sstates$, full revelation followed by the algorithm in part~(i), with accuracy $3\varepsilon/4$, gives a joint signaling and contract scheme whose principal utility is at least
\[
\frac{\OPT(\prior)}{4/(3\gamma_\star)+\varepsilon},
\qquad
\frac{4}{3\gamma_\star}=3.0803111140\ldots.
\]
\end{enumerate}
The algorithm performs $n^{\poly(1/\varepsilon)}$ arithmetic operations and value-oracle queries, per realized state in part~(ii).
\end{theorem}

The algorithm and the proof of \Cref{thm:computation} appear in \Cref{app:algorithms}. The complexity bound is polynomial for each fixed $\varepsilon$; for an explicitly listed finite state space, the entire contract table can be computed by running the algorithm once for each state.

%% Transparency with Complementarity
\section{Transparency with Complementarity}
%\label{sec:among_workers}
\label{sec:complements}

In this section, we show that, for complementary tasks (i.e., the success probability is a supermodular function), the loss from transparency can be unbounded. 
This leads to a sharp contrast with the tight $\frac43$-approximation for the substitutable case in \cref{sec:multi-agents}.

\begin{theorem}\label{thm:supermodular}
When the reward function is supermodular, there exists an instance in which the principal receives zero payoff by fully revealing the state, while revealing no information generates a strictly positive payoff. 
\end{theorem}

\begin{proof}
Consider a setting with two agents and two states that occur with equal probabilities. 
To simplify the presentation, we will construct the example with expected rewards that are $> 1$.
In the following example, dividing every reward and cost below by $8$ yields rewards in $[0,1]$, leaves all incentive shares unchanged, and divides every principal payoff by $8$.

Specifically, the costs of the agents for exerting effort are $c_1=2, c_2=4$. In both states of the world, the expected reward is $0$ when neither of the two agents exerts effort.
\begin{itemize}
    \item In state $0$, the expected reward %return 
    is $2$ when only one of the agents exerts effort and it is $8$ when both agents exert effort. 
    \item In state 1, the expected reward is $\frac{20}{21}$ when only agent 1 exerts effort, it is $0$ when only agent $2$ exerts effort, and it is $\frac{20}{3}$ when both agents exert effort. 
\end{itemize}
Suppose the state is perfectly revealed to the agents. 
\begin{itemize}
\item In state $0$, the principal can either offer a contract $(\alpha_1,\alpha_2) = (1,0)$ to incentivize agent $1$ to exert effort, or a contract $(\alpha_1,\alpha_2) =  (\frac{1}{3},\frac{2}{3})$ to incentivize both agents to exert effort. Agent $2$ cannot be incentivized to exert effort by himself with a linear contract less than $1$. In all cases, the principal's payoff is $0$. 
\item In state $1$, the principal cannot incentivize any of the agents to exert effort on their own with a linear contract $(\alpha_1,\alpha_2)$ such that $\alpha_1 + \alpha_2 \leq 1$. To incentivize both of them to exert effort, 
the principal needs to offer a contract $(\alpha_1,\alpha_2) = (\frac{3}{10},\frac{7}{10})$. In all cases, the principal's payoff is $0$. 
\end{itemize}
In summary, the principal cannot guarantee a positive payoff by fully revealing the states. 

However, the principal can achieve a strictly positive payoff by providing no information to the agent and offering a contract $(\alpha_1,\alpha_2) = (\frac{6}{19},\frac{28}{41})$. 
Given this contract, there exists a pure Nash equilibrium in which both agents exert effort. The expected payoff of the principal is 
\begin{equation*}
\left(1-\frac{6}{19}-\frac{28}{41}\right)\cdot\frac{22}{3} = \frac{22}{2337} > 0.\qedhere
\end{equation*}
\end{proof}

%% Section: Conclusion
\section{Conclusion}
\label{sec:conclude}
In this paper, we quantify the price of transparency in team contracts. We show that full transparency achieves a tight $\frac43$-approximation to the optimal joint information disclosure and contracting scheme if agents are substitutable (i.e., the production function is submodular).  In contrast, the loss is unbounded in the worst case if agents are complementary. Our pruning approach also yields deterministic polynomial-time approximation algorithms with ratios $2.311$ for the known-state contract problem and $3.081$ for the joint information and contracting problem. 

Following our model, there are many directions worthy of further investigation. 
For example, in the complementary case, since full transparency may incur an unbounded loss, it would be interesting and useful to find a simple information structure that provides a constant approximation guarantee to the optimal one. 
Moreover, our paper focuses on public signals. In general, there may be 
benefits to keeping the workers asymmetrically informed through private signals. 
It would be interesting to quantify the gap between public and private signals for both the substitutable and complementary cases. 
Finally, our paper focuses on environments with binary actions. In \cref{apx:multi-action}, we provide a generalization to multiple actions and show that even in the single-agent case, full transparency leads to a $\Theta(n)$-approximation to the optimal joint design where $n$ is the number of actions. 
It is an open question whether constant approximations are possible in multi-action settings if we impose additional structure on the contracting environments. 

\section*{AI Methodology}
The initial set of results in this paper was entirely human-generated. This includes the All-or-Nothing Lemma, as well as a pruning-based argument that lead to a (non-tight) constant-factor approximation guarantee for submodular functions. The unbounded gap construction for supermodular functions was also part of this initial seed. 

The authors subsequently used state-of-the-art LLMs (such as ChatGPT Astra, Claude Opus, and Gemini Argon) to assist with tightening the bound for submodular functions (see Section~\ref{sec:multi-agents}), and with generalizing the guarantee from pure to mixed equilibria (Appendix~\ref{app:BeyondPureNash}). %More details can be found in [Section Z]. 
The models also assisted the proofs that translate the Potential Lemma into an improved poly-time approximation algorithm (Appendix~\ref{app:algorithms}). We provide LLM-generated and verifed Lean certificates at \url{http://github.com/wuzhuangtai00/transparency-lean}.

The authors re-organized and refactored all AI-assisted proofs to ensure readability and carefully checked correctness.

\section*{Acknowledgments}

This work was supported by the European Research Council (ERC) under the European Union's Horizon 2020 research and innovation program (grant agreement No.~101077862, project ALGOCONTRACT), by the Israel Science Foundation (grant No.~3331/24), by the NSF-BSF (grant No.~2021680), and by a Google Research Scholar Award. Y.~Li thanks the NUS Start-up Grant for financial support. J. Wu was supported by the NSF Awards CCF-1942583 (CAREER), CCF-2342642, and a CADMY Research Fellowship.

\newpage

%% References
\bibliographystyle{apalike}
\bibliography{ref}
\newpage

%% Appendices
\appendix

%% Section: Extension to Mixed Nash Equilibria
\section{Extension to Mixed Nash Equilibria}\label{app:BeyondPureNash}

We show that all our bounds on the price of transparency hold with the same approximation factors when the principal may select a mixed Nash equilibrium after every public signal. In particular, the tight $\frac43$-approximation for submodular rewards continues to hold when mixed equilibria are allowed under both full transparency and the optimal information scheme. The reduction below encodes a mixed strategy profile as a pure equilibrium of an auxiliary instance, while allowing the auxiliary full-transparency solutions to be implemented as mixed equilibria in the original instance.

\paragraph{Mixed equilibria and benchmarks.}
Let $x\in[0,1]^n$ denote a mixed strategy profile, and let $R_x$ be a random set containing each agent $i$ independently with probability $x_i$. For each state $\omega$, define the multilinear extension
\[
F_\omega(x):=\expect{f_\omega(R_x)},\qquad
F_\mu(x):=\expect[\omega\sim\mu]{F_\omega(x)}.
\]
Write $R_{x,-i}:=R_x\setminus\{i\}$ and
\[
\partial_i F_\mu(x)=\expect{f_\mu(i\mid R_{x,-i})}
\]
for agent $i$'s expected marginal contribution. For a stateless reward function $f$, we write $F$ for its multilinear extension. The profile $x$ is a mixed Nash equilibrium of contract $\alpha$ at posterior $\mu$ if, for every $i$,
\begin{equation}\label{eq:mixed-best-responses}
\alpha_i\partial_i F_\mu(x)
\begin{cases}
\le c_i,&x_i=0,\\
=c_i,&0<x_i<1,\\
\ge c_i,&x_i=1.
\end{cases}
\end{equation}
The principal's utility is $(1-\sum_i\alpha_i)F_\mu(x)$. Let $\OPT_{\mathrm{mix}}(f,c)$ be its maximum over contracts and mixed Nash equilibria in a stateless instance. We use $\OPT_{\mathrm{mix}}(\prior)$, $U^F_{P,\mathrm{mix}}(\prior)$, and $U^N_{P,\mathrm{mix}}(\prior)$ for the corresponding joint-design, full-transparency, and no-disclosure benchmarks. Thus
\[
U^F_{P,\mathrm{mix}}(\prior)
=\expect[\omega\sim\prior]{\OPT_{\mathrm{mix}}(f_\omega,c)},\qquad
U^N_{P,\mathrm{mix}}(\prior)=\OPT_{\mathrm{mix}}(f_\prior,c).
\]
For these maximization problems, we may restrict to contracts with total share at most one: larger total shares give nonpositive principal utility, whereas the zero-payment contract admits an equilibrium with nonnegative principal utility.

\paragraph{The reduction.}
For $T\subseteq[n]$, let $x^T$ denote the vector whose $i$th coordinate is $x_i$ if $i\in T$ and zero otherwise.

\begin{lemma}[Mixing reduction]\label{lem:mixing-reduction}
Fix a contract setting $(\prior,\{f_\omega\},c)$ and a profile $x\in[0,1]^n$. Define an auxiliary setting with the same states and prior by
\begin{equation}\label{eq:mixed-auxiliary-instance}
\widetilde f_\omega(T):=F_\omega(x^T)
=\expect{f_\omega(T\cap R_x)},\qquad
\widetilde c_i:=x_i c_i.
\end{equation}
Then the following hold:
\begin{enumerate}[(i)]
\item Each $\widetilde f_\omega$ is monotone and takes values in $[0,1]$. If $f_\omega$ is submodular, supermodular, or additive, then $\widetilde f_\omega$ has the same property.
\item If $x$ is a mixed Nash equilibrium of a contract $\alpha$ at posterior $\mu$ in the original setting, then $[n]$ is a pure Nash equilibrium of the same contract at posterior $\mu$ in the auxiliary setting, with the same principal utility.
\item In every state $\omega$, any auxiliary team $T$ with a feasible minimum-payment contract $\beta$ can be implemented in the original setting by the contract $\beta$ and the mixed Nash equilibrium $x^T$, with the same principal utility. Consequently,
\begin{equation}\label{eq:mixed-statewise-comparison}
\OPT_{\mathrm{mix}}(f_\omega,c)
\ge\OPT(\widetilde f_\omega,\widetilde c).
\end{equation}
\end{enumerate}
\end{lemma}

\begin{proof}
For each fixed set $R$, the function $T\mapsto f_\omega(T\cap R)$ preserves monotonicity and each of the stated structural properties of $f_\omega$. Taking expectations proves~(i).

For~(ii), the gain of auxiliary agent $i$ from working when all other auxiliary agents work is
\[
\alpha_i\bigl(\widetilde f_\mu([n])-\widetilde f_\mu([n]\setminus\{i\})\bigr)-\widetilde c_i
=x_i\bigl(\alpha_i\partial_i F_\mu(x)-c_i\bigr)\ge0.
\]
The inequality follows from~\eqref{eq:mixed-best-responses}, including when $x_i=0$. Since $\widetilde f_\mu([n])=F_\mu(x)$, the principal's utility is unchanged.

For~(iii), fix $\omega$ and an auxiliary team $T$ with its minimum-payment contract $\beta$, and put $y=x^T$. For every $i\in T$ with $x_i c_i>0$, independence gives
\[
\widetilde f_\omega(i\mid T\setminus\{i\})
=x_i\partial_i F_\omega(y),
\]
and hence
\begin{equation}\label{eq:mixed-implementation-share}
\beta_i
=\frac{x_i c_i}{\widetilde f_\omega(i\mid T\setminus\{i\})}
=\frac{c_i}{\partial_i F_\omega(y)}.
\end{equation}
The denominator is positive because $T$ is feasible and $x_i c_i>0$. Thus each such agent is indifferent between effort and shirking, and its prescribed probability $y_i$ is a best response. An agent with $x_i c_i=0$ receives zero share: if $x_i=0$, shirking is a best response in the original game; if $c_i=0$, every mixing probability is a best response. Agents outside $T$ also receive zero shares and can shirk. Therefore $y$ is a mixed Nash equilibrium. Its success probability is $F_\omega(y)=\widetilde f_\omega(T)$, so it gives the principal the same utility. Every auxiliary pure-equilibrium optimum can be attained using a minimum-payment contract, which proves~\eqref{eq:mixed-statewise-comparison}.
\end{proof}

\paragraph{Submodular rewards.}
We now apply the reduction to the transparency guarantee. The approximation factors compare the best mixed equilibria under the two information benchmarks; there is no additional loss from passing through the auxiliary pure-equilibrium instance.

\begin{theorem}\label{thm:mixed-transparency}
If $f_\omega$ is submodular for every state $\omega$, then
\[
U^F_{P,\mathrm{mix}}(\prior)
\ge\frac34\,\OPT_{\mathrm{mix}}(\prior).
\]
The factor $\frac43$ is tight, even when every $f_\omega$ is additive.
\end{theorem}

\begin{proof}
Fix any contract $\alpha$ and mixed Nash equilibrium $x$ under no disclosure. Form the auxiliary instance in~\eqref{eq:mixed-auxiliary-instance}, and let $\widetilde U^F_P$ and $\widetilde U^N_P$ denote its pure-equilibrium full-transparency and no-disclosure benchmarks. By \Cref{lem:mixing-reduction}(ii),
\[
\widetilde U^N_P(\prior)
\ge\Bigl(1-\sum_i\alpha_i\Bigr)F_\prior(x).
\]
Applying \Cref{thm:submodular} to the auxiliary submodular instance and then \Cref{lem:mixing-reduction}(iii) in each state gives
\begin{align*}
U^F_{P,\mathrm{mix}}(\prior)
&\ge\widetilde U^F_P(\prior)
\ge\frac34\,\widetilde U^N_P(\prior)\\
&\ge\frac34\Bigl(1-\sum_i\alpha_i\Bigr)F_\prior(x).
\end{align*}
Maximizing over $\alpha$ and $x$ proves
$U^F_{P,\mathrm{mix}}(\prior)\ge\frac34 U^N_{P,\mathrm{mix}}(\prior)$ for every prior. The All-or-Nothing Lemma (\Cref{lem:approximation_extension}) applies unchanged: its proof uses only the posterior interpretation and the law of iterated expectations. Explicitly, for any public signaling scheme $\pi$ with induced posterior distribution $\pi\given\prior$,
\[
U^F_{P,\mathrm{mix}}(\prior)
=\expect[\mu\sim\pi\given\prior]{U^F_{P,\mathrm{mix}}(\mu)}
\ge\frac34\expect[\mu\sim\pi\given\prior]{U^N_{P,\mathrm{mix}}(\mu)}.
\]
The last expectation upper-bounds the utility of any contract scheme and mixed-equilibrium selection associated with $\pi$. Optimizing over $\pi$ proves the upper bound.

For tightness, suppose $f$ is additive. Under any contract, an agent's incentive to work is independent of the other agents' actions. Replacing every positive mixing probability by one therefore preserves equilibrium and weakly increases the success probability. For a contract with total share at most one, this weakly increases principal utility. Thus the optimal pure- and mixed-equilibrium utilities coincide for additive rewards, both in every revealed state and at every posterior. The additive instances of \Cref{prop:submodular-tight} therefore retain their ratios, which approach $\frac43$.
\end{proof}

\paragraph{The other transparency bounds.}
For a single agent, a mixed best response is supported on pure best responses. The principal can select the pure action in this support giving her the highest utility, which weakly improves on the mixture. This applies at every posterior and for every contract, including with multiple actions. Hence the tight $\frac98$ guarantee of \Cref{thm:approx_fully_revealing_binary} and the $\Theta(n)$ bound of \Cref{thm:multi_action} are unchanged.

For supermodular rewards and binary actions, the optimal stateless principal utility also coincides under pure and mixed equilibria. To see this, fix a mixed equilibrium $x$ under a contract $\alpha$ with total share at most one, and let $S=\{i:x_i>0\}$. Supermodularity and~\eqref{eq:mixed-best-responses} imply, for every $i\in S$,
\[
\alpha_i f(i\mid S\setminus\{i\})
\ge\alpha_i\partial_i F(x)\ge c_i.
\]
Keep the shares of agents in $S$ and set all other shares to zero. Then $S$ is a pure Nash equilibrium, the total share weakly decreases, and $f(S)\ge F(x)$, so principal utility weakly increases. Since posterior averages of supermodular functions are supermodular, this equality holds at every posterior as well as in every revealed state. Thus the unbounded loss from full transparency in \Cref{thm:supermodular} persists when mixed equilibria are allowed under both information benchmarks.

%% Section: Tight Analysis of the Single-Agent Setting
\section{Tight Analysis of the Single-Agent Case}
\label{sec:single-agent}

In this section we focus on the important special case of a single agent. We use \Cref{lem:approximation_extension} to characterize a tight approximation guarantee for full transparency in single-agent environments, showing it is a small constant. The resulting ratio of $\frac98$ is better than the tight ratio of $\frac43$ that we establish for general (multi-agent) settings %with submodular rewards 
(\Cref{sec:multi-agents}).

In single-agent environments, the only quantities that matter are the agent's cost $c$ for exerting effort, and the probabilities of success conditional on the agent exerting effort or not exerting effort, 
which we denote by $p_1$ and $p_0$ respectively (where the state of the world $\omega$ is clear from context). More formally, when $n=1$ every success function $f_\omega$ is fully specified by $f_\omega([n])=p_1$ and $f_\omega(\emptyset)=p_0$. By monotonicity, $p_0\le p_1$. Thus, in this section, a contract setting is described by $(\prior,\{(p^\omega_0,p^\omega_1)\}_{\omega\in\Omega},c)$.

\begin{theorem}
\label{thm:approx_fully_revealing_binary}
For every single-agent contract setting with states, it holds that full revelation is a $\frac{9}{8}$-approximation to the optimal information structure, and the factor $\frac{9}{8}$ is tight. Moreover, full revelation is optimal if $p_0=0$ in every state of the world. %with probability $1$.
\end{theorem}

The suffix of the theorem is not hard to see: If $p_0=0$ in every state then an outcome of success reveals with certainty that the agent exerted effort. Under full revelation, a payment of $c/p_1$ upon success when $p_1>c$, and zero payments otherwise, therefore extracts the full surplus.
We devote the remainder of the section to proving the $9/8$-approximation. If $c=0$, both no and full revelation give $\expect{p_1}$; if $c\ge1$, both give $\expect{p_0}$. We henceforth assume $0<c<1$ and restrict ratios to $U^F_P(\prior)>0$. If $U^F_P(\prior)=0$, then $U^N_P(\prior)=0$ as well. By \Cref{lem:approximation_extension} we can focus on the approximation ratio between no and full revelation, i.e., on 
$$
\sup_{\prior}\{U^N_P(\prior)/U^F_P(\prior)\}.
$$

\subsection{Narrowing Down the Search Space}

%\subsection{Stateless Single-Agent Contracts}
We start with two observations concerning the optimal principal's utility in a stateless contract setting (for a fixed state $\omega$). We use these observations below to narrow down the space of worst-case distributions we need to consider. 
%The proofs of the observations appear in \Cref{apx:single-agent}. 

\paragraph{Stateless Single-Agent Contracts.}

\begin{observation}\label{obs:indifference-point} 
Fix state $\omega$, let $p_0$ be the success probability if the agent does not exert effort, and $p_1$ be the success probability upon effort. The principal's optimal utility from incentivizing no effort is $p_0$, and for effort it is $p_1 - p_1 \cdot c/(p_1-p_0)$ provided $p_1-p_0\ge c$; otherwise effort cannot be induced by a share in $[0,1]$. Let
\begin{align*}
\lp(p_0) := p_0 + \frac{c + \sqrt{c^2+4cp_0}}{2}.
\end{align*}
Then the principal prefers to incentivize the agent not to exert effort if $p_1 \leq \ell(p_0)$, she is indifferent between incentivizing no effort and effort if $p_1 = \ell(p_0)$, and she prefers to incentivize the agent to exert effort if $p_1 \geq \ell(p_0)$.
\end{observation}

\begin{proof} %[Proof of \Cref{obs:indifference-point}]
Since the cost of no effort is normalized to zero, the principal can incentivize the agent to not exert effort with zero payment. So the principal's optimal utility from no effort is $p_0$. If $p_1=p_0$, effort is strictly worse for the agent since $c>0$, so the principal obtains $p_0$. For $p_1>p_0$, to incentivize effort, the principal must offer a payment of $t$ so that $p_1 \cdot t - c = p_0 \cdot t$. Solving for $t$, this yields $t = c / (p_1-p_0).$ This share is feasible exactly when $p_1-p_0\ge c$. In that case, the principal's optimal utility from incentivizing effort is $
p_1 - p_1 \cdot c/(p_1-p_0).$ Observing that $p_0$ is a constant when viewed as a function of $p_1$, while $p_1 - p_1 \cdot c/(p_1-p_0)$ is increasing in $p_1$, 
we obtain the claim by equating the respective utilities, $p_0$ and $p_1 - p_1 \cdot c/(p_1-p_0)$, and solving for $p_1$.
\end{proof}

\begin{figure}[t]
    \centering
    % Requires: \usepackage{pgfplots}
    % Optional: \pgfplotsset{compat=1.18}
    \begin{tikzpicture}
    % --- tweakables (smaller P0 and \ell(P0); smaller V(P0,1)) ---
    \pgfmathsetmacro{\Pzero}{0.20}   % fixed p_0 in [0,1)
    \pgfmathsetmacro{\ellx}{0.45}    % kink location \ell(p_0) in (p_0,1]
    \pgfmathsetmacro{\kappa}{12.0}   % curvature (>0): bigger = more concave
    \pgfmathsetmacro{\amp}{0.35}     % amplitude after kink
    
    % Value at x=1 for the illustrative curve (purely for tick labeling)
    \pgfmathsetmacro{\Vone}{ \Pzero + \amp * (1 - exp(-\kappa*(1-\ellx))) }
    
    % Choose a compact y-maximum explicitly (shorter y axis).
    \pgfmathsetmacro{\ymax}{ \Pzero + 1.05*\amp }  % small headroom above V(P0,1)
    
    \begin{axis}[
      width=8.8cm,
      axis equal image,            % same scale on x and y axes
      xmin=0, xmax=1.02,
      ymin=0, ymax=\ymax,          % explicitly short y-axis
      axis lines=left,
      xlabel={$p_1$},
      ylabel={},
      % place labels near axis ends; y-label horizontal
      xlabel style={at={(axis description cs:1.02,0)}, anchor=west},
      ylabel style={at={(axis description cs:0,1.02)}, anchor=south, rotate=0},
      xtick={0,\Pzero,\ellx,1},
      xticklabels={{$0$},{$p_0$},{$\ell(p_0)$},{$1$}},   % <-- double braces
      ytick={0,\Pzero,\Vone},
      yticklabels={{$0$},{$p_0$},{$U^N_P(\prior)$}},          % <-- double braces (comma-safe)
      samples=300,
      clip=false
    ]
    
    % Flat segment: V = P0 on [p_0, \ell(p_0)]
    \addplot[very thick, blue]
      coordinates {(\Pzero,\Pzero) (\ellx,\Pzero)};
    
    % Concave, increasing (illustrative):
    % V = P0 + amp * (1 - e^{-kappa (x - ell)})
    \addplot[very thick, blue, domain=\ellx:1]
      ({x},{ \Pzero + \amp * (1 - exp(-\kappa*(x-\ellx))) });
    
    % Kink indicator
    \addplot[dashed, gray] coordinates {(\ellx,0) (\ellx,\ymax)};
    \addplot[only marks, mark=*, mark size=1.7pt, blue] coordinates {(\ellx,\Pzero)};
    
    % Region labels (optional)
    \node[anchor=north west] at (axis cs:\Pzero+0.06,\Pzero) {flat};
    \node[anchor=south west] at (axis cs:\ellx+0.12,\Pzero+0.17) {concave};
    
    \end{axis}
    \end{tikzpicture}

    \caption{
    Consider a single-agent contract setting with a single state $\omega$ (the prior is a point-mass
    $\prior=\delta_{\omega}$). 
    Fix the success probability if the agent does not exert effort to be $p^{\omega}_0=p_0$. Vary the success probability upon effort $p^{\omega}_1=p_1$ between $p_0$ and $1$. 
    The optimal payoff of the principal $U^N_P(\prior)$ as a function of $p_1$ is flat then concave. The point $\ell(p_0)$ is defined in \Cref{obs:indifference-point}.}
    \label{fig:vn}
\end{figure}
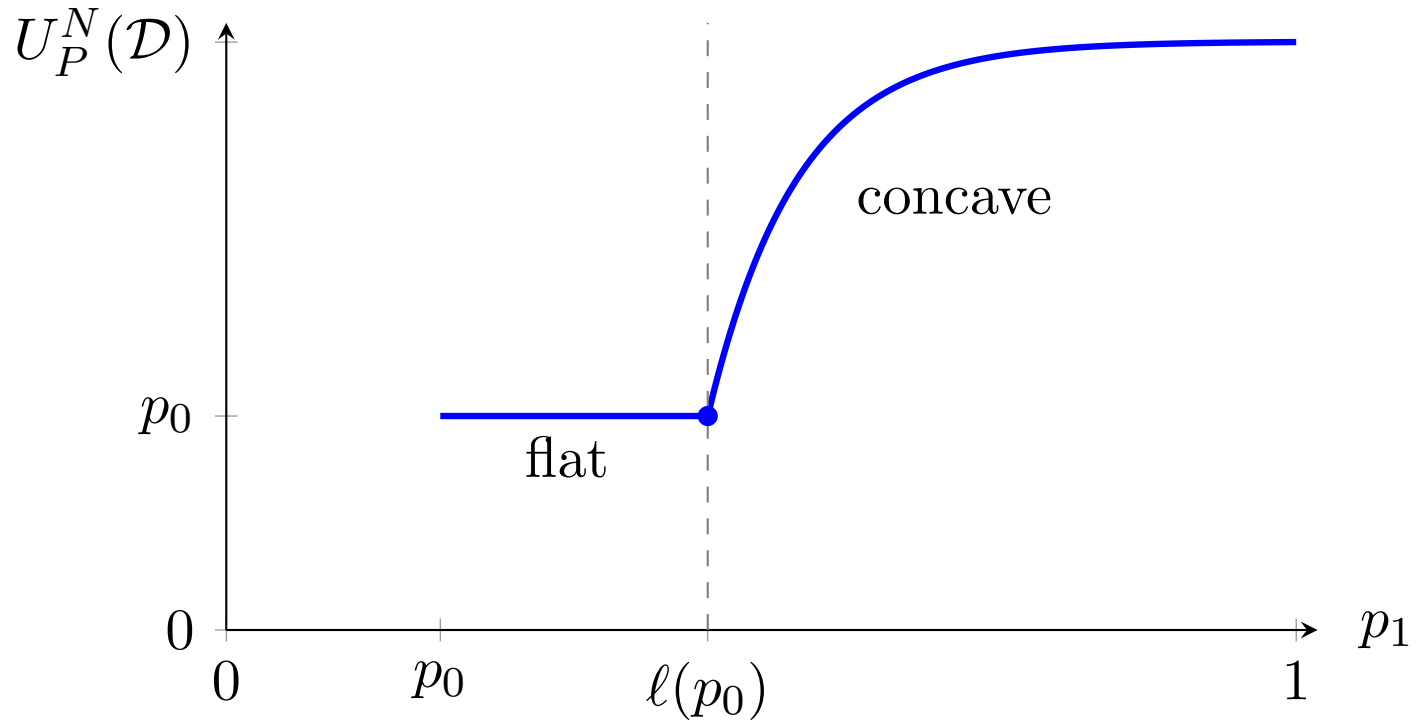

\Cref{fig:vn} illustrates the following observation: 

\begin{observation}\label{obs:optimal-utility}
    The optimal payoff of the principal for a fixed state $\omega$, as a function of $p_1$ when fixing $p_0$, is given by $\max\{p_0,p_1-p_1\cdot c/(p_1-p_0)\}$ for $p_1>p_0$, and equals $p_0$ when $p_1=p_0$. This is a constant function for $p_1 \leq \lp(p_0)$ and a concave function for $p_1 \geq \lp(p_0)$.
\end{observation}

\begin{proof}%[Proof of \Cref{obs:optimal-utility}]
By \Cref{obs:indifference-point}, no effort gives $p_0$, and feasible effort gives $p_1-p_1c/(p_1-p_0)$. For $0<p_1-p_0<c$, this latter expression is negative and is dominated by $p_0$; for $p_1=p_0$, the value is $p_0$.

Hence the optimal principal's utility, when viewed as a function of $p_1$, is $p_0$ for $p_1 \leq \ell(p_0)$, while it is $p_1 - p_1 \cdot c/(p_1-p_0)$ for $p_1 \geq \ell(p_0)$. To see that the latter is a concave function in $p_1$, observe that the second derivative of $p_1 - p_1 \cdot c/(p_1-p_0)$ in $p_1$ is $- (2 c p_0)/(p_1 - p_0)^3$, which is non-positive (since $p_1 \geq p_0$ by monotonicity).
\end{proof}

\paragraph{Distributions Over Two States Suffice.}

The next lemma uses the observations about stateless contracting (\Cref{obs:indifference-point} and \Cref{obs:optimal-utility}) to narrow down the space of possible worst-case distributions. It shows that, in order to bound the worst-case gap between no and full revelation, it is sufficient to consider distributions over two states of the world.
In one supported state, there is no hidden action, as effort leads deterministically to success, and no effort always leads to failure; in the other state, the principal is indifferent to the agent's effort.  

\begin{lemma}\label{lem:worst_case_single}
Fix $c\in(0,1)$. Among all single-agent contract settings $(\prior,\{(p^\omega_0,p^\omega_1)\},c)$, 
the supremum of the approximation ratio
$\sup_{\prior}\{U^N_P(\prior)/U^F_P(\prior)\}$
is achieved by a two-point prior $\prior^\star$ supported on $\omega$ and $\omega'$, where
$(p_0^\omega,p_1^\omega)=(0,1)$ and $(p_0^{\omega'},p_1^{\omega'})=(p_0,\lp(p_0))$
for some $p_0\in[0,1]$ with $\lp(p_0)\le 1$.
\end{lemma}

\begin{proof}%[Proof of \cref{lem:worst_case_single}]
When seeking a ratio above one, we may first assume $p_0\le r:=\ell^{-1}(1)=1-\sqrt c$ in every supported state.\footnote{For $p_0>r$, first raise $p_1$ to $1$, leaving the full-revelation payoff $p_0$ unchanged and weakly increasing the no-revelation payoff. Split $(p_0,1)$ into $(r,1)$ and $(1,1)$, preserving both means and the full-revelation payoff. To discard $(1,1)$, let the remaining conditional prior have means $(a,b)$ and no- and full-revelation payoffs $N,F$. Comparing the two payoff branches in \Cref{obs:optimal-utility} gives $U_P^N(1-\lambda+\lambda a,1-\lambda+\lambda b)\le1-\lambda+\lambda U_P^N(a,b)$ for $\lambda\in[0,1]$. Adding mass $1-\lambda$ on $(1,1)$ therefore gives a ratio at most $(1-\lambda+\lambda N)/(1-\lambda+\lambda F)\le\max\{1,N/F\}$. If $F=0$, then $N=0$ and the ratio is at most one.}
First, since $U^N_P(p_0,p_1)$ is constant in $p_1$ for $p_1 \leq \lp(p_0)$, in the worst case, for any realized pair of $(p_0, p_1)$, we must have 
$p_1 \geq \lp(p_0)$. 
Otherwise, by \Cref{obs:indifference-point}, we can increase the realization of $p_1$ for that realization, which weakly increases the payoff from revealing no information, while the payoff from revealing full information remains the same. 

Moreover, since $U^N_P(p_0,p_1)$ is concave in $p_1$ for $p_1 \geq \lp(p_0)$, 
to minimize the payoff from revealing full information, 
conditional on $p_0$, it is optimal to induce a distribution over posterior support on $\{\lp(p_0), 1\}$. 
Therefore, the support of the distribution has two categories: 
(1) $(p_0,1)$; and (2) $(p_0,\lp(p_0))$.
\begin{itemize}
\item Supports with the form $(p_0,1)$. First, observe that the principal's payoff $U^N_P(p_0,1)$ is concave in $p_0$ for the range of $p_0\in[0,\lp^{-1}(1)]$. Therefore, to minimize the principal's payoff from fully revealing, 
it is sufficient to consider distributions with a binary support of
$(0,1)$ and $(\lp^{-1}(1),1)$. 
Moreover, note that the latter can be combined with the case where the state takes the form of $(p_0,\lp(p_0))$ with $p_0 = \lp^{-1}(1)$.

\item Supports with the form $(p_0,\lp(p_0))$. In this case, note that the cutoff function $\lp(p_0)$ is concave in $p_0$. 
Letting $(\hat{\mean}_0,\hat{\mean}_1)$ be the marginal means for states $(p_0,\lp(p_0))$,
Jensen's inequality implies that $\lp(\hat{\mean}_0) \geq \hat{\mean}_1$.
Therefore, \Cref{obs:optimal-utility} implies that the principal's optimal payoff remains unchanged as $\hat{\mean}_0$ from pooling states $(p_0,\lp(p_0))$ by incentivizing the agent not to exert effort. Finally, raise the pooled state's effort probability to $\ell(\hat{\mean}_0)\le1$, preserving the full-revelation payoff and weakly increasing the no-revelation payoff.
\end{itemize}

Combining the arguments, the worst case distribution has a binary support on $(0,1)$ and $(p_0,\lp(p_0))$.
\end{proof}

\paragraph{The Worst-Case Approximation Ratio.}

Fix any such distribution $\mathcal{D}^\star$ as in \Cref{lem:worst_case_single}. Let $q$ denote the probability of $\omega$, and let $m_0  = m_0(\mathcal{D}^\star) = (1-q)p_0$ and $m_1 = m_1(\mathcal{D}^\star) = q + (1-q)\ell(p_0)$ denote the mean of the success probabilities under no effort and under effort, respectively. 

Using this notation, the principal's optimal utility under full revelation is
\begin{align*}
U^F_P(\prior^\star)
&= q\cdot U^N_P(0,1)\;+\;(1-q)\cdot U^N_P\bigl(p_0,\lp(p_0)\bigr)
= q(1-c)\;+\;(1-q)\,p_0.
\end{align*}

The principal's optimal utility under no revelation only depends on the means $m_0$ and $m_1$. Using $U_P^N(m_0,m_1)$ to denote the principal's optimal utility from no revelation under any distribution with these means, an argument analogous to \Cref{obs:optimal-utility} shows
\[
U_P^N(\mathcal{D}^\star) = U_P^N(m_0,m_1) = \max\left\{m_0,\left(1-\frac{c}{m_1-m_0}\right)\cdot m_1\right\}.
\]

Note that since we're interested in the worst-case approximation ratio, and so in enlarging the ratio between $U^N_P$ and $U^F_P$ as much as possible, we only need to consider the case $\mean_1 > \lp(\mean_0)$. Otherwise, the principal does not benefit from incentivizing the agent to exert effort when providing no information, and
can attain a weakly higher payoff under full information by simply providing a zero contract in all states. 

\subsection{A Mathematical Program}

Based on the previous discussion, we can capture the worst-case approximation ratio through a mathematical program. This program maximizes the ratio $U_P^N(\mathcal{D}^\star)/U_P^F(\mathcal{D}^\star)$ over all $\mathcal{D}^\star$ as parameterized by $q, p_0, c$.
To solve that program, we introduce a re-parameterization, which will be useful for the analysis. 
Specifically, given cost $c$, we introduce two nonnegative parameters $(X,k)$ and let  
\[
q:=kc \quad \text{and} \quad \Delta_B \;:=\; \lp(p_0)-p_0 \;=\; Xc.
\]
By algebraic calculations, this implies that $p_0 = cX(X-1)$ and $\lp(p_0)=p_0+\Delta_B=cX^2$.
Moreover, 
\[
\mean_0 \;=\; (1-q)p_0 \;=\; c(1-kc)X(X-1)\quad \text{and} \quad
\mean_1 \;=\; q\cdot 1+(1-q)\lp(p_0) \;=\; c\bigl[k+X^2-kX^2c\bigr].
\]
In this instance, feasibility ($q\in [0,1], \lp(p_0)\in[0,1]$) requires 
\[
c \;\le\; c_{\max}(k,X) \;:=\; \min\!\Bigl\{\frac{1}{k},\ \frac{1}{X^2}\Bigr\}.
\]
Here $1/k=+\infty$ when $k=0$. All ratio formulas below are restricted to positive full-revelation payoff, equivalently $k(1-c)+(1-kc)X(X-1)>0$; in particular, the zero-payoff case $k=0$, $X=1$ is excluded.
Therefore, the worst approximation reduces to the following optimization program.
\begin{equation}\label{eq:Rkx}
\sup_{\prior}\frac{U^N_P(\prior)}{U^F_P(\prior)}
= \sup_{k\geq 0, X\geq 1, c\in\bigl(0,\,c_{\max}(k,X)\bigr]} R(c;k,X) \;:=\; \frac{\mean_1\!\left(1-\frac{c}{\mean_1-\mean_0}\right)}{c\Bigl[k(1-c)+(1-kc)X(X-1)\Bigr]}.
\end{equation}

We solve the mathematical program by first showing a monotonicity property, implying that the worst case approximation occurs in the limit when the cost $c\to 0$. 
Then, we determine the supremum by calculating the exact limit value as $c\to 0$.
\begin{lemma}\label{lem:R-monotone}
For any fixed $k\geq 0,X\geq1$, the approximation ratio
$R(c;k,X)$ is  non-increasing in $c$ for $c\in(0,c_{\max}(k,X)]$ wherever its denominator is positive.
\end{lemma}

The proof of \cref{lem:R-monotone} consists primarily of algebraic calculations and appears in \cref{sub:algebra}.

\begin{lemma}\label{lem:limit_approx}
It holds that
\[
\sup_{k,X}\ \lim_{c\to 0} R(c;k,X) \ =\ \frac{9}{8}.
\]
\end{lemma}

\begin{proof}[Proof of \cref{lem:limit_approx}]
By \Cref{lem:R-monotone}, given any $(k,X)$, the function $R(c;k,X)$ attains its supremum on the feasible interval at the limit $c\to 0$.

From the definition in \eqref{eq:Rkx}, we have
\[
R_0(k,X) \;:=\; \lim_{c\to 0} R(c;k,X)
\;=\; 1 + \frac{k(X-1)}{(k+X)\,(k+X^2-X)}.
\]
For $X=1$ and $k>0$, $R_0(k,1)=1$. Maximizing in $k>0$ for fixed $X>1$ yields the unique interior optimizer
$k^*(X)=X\sqrt{X-1}$. Substituting $k^*(X)$ into $R_0$ and maximizing in $X\ge 1$
gives the peak at $X=2$ with $k^*(2)=2$, and
\[
R_0(2,2) \;=\; 1 + \frac{2}{16} \;=\; \frac{9}{8}.
\]
Therefore, the worst-case ratio is $\frac{9}{8}$, attained in the limit $c\to 0$.
\end{proof}

We are now ready to prove the main theorem of this section.

\begin{proof}[Proof of \cref{thm:approx_fully_revealing_binary}]
The proof of the claimed (tight) worst-case approximation guarantee follows by combining \Cref{lem:approximation_extension} with Lemmas \ref{lem:worst_case_single}-\ref{lem:limit_approx}.
\end{proof}

\subsection{Proof of Lemma~\ref{lem:R-monotone}}
\label{sub:algebra}

\begin{proof}[Proof of \cref{lem:R-monotone}]
We first recall the necessary definitions and constraints. 
Let 
\[
\Delta_P \;:=\; \mean_1-\mean_0 \;=\; c\bigl(k+X-kXc\bigr)\;=\; c\,H(c),
\quad\text{where } H(c):=k+X-kXc.
\]
In the relevant region $\Delta_P\ge c$ (i.e., no information induces effort), we have
\[
U^N_P(\prior) \;=\; \mean_1\!\left(1-\frac{c}{\Delta_P}\right)
\;=\; \mean_1\!\left(1-\frac{1}{H(c)}\right).
\]
The approximation ratio is then given by:
\begin{equation}
R(c;k,X) = \frac{U^N_P(\prior)}{U^F_P(\prior)}
= \frac{\mean_1\!\left(1-\frac{1}{H(c)}\right)}{c\Bigl[k(1-c)+(1-kc)b\Bigr]},
\end{equation}
where $b=X(X-1)$ and $\mean_1/c = k+X^2-kX^2c$.
The constraints are $X\ge 1$ (since $p_0=cb \ge 0$) and $k\ge 0$. The feasible range for $c$ is $(0, c_{\max}]$, where $c_{\max}(k,X) = \min\{1/k, 1/X^2\}$.

We utilize the substitution $a=1-kc$. As $c$ ranges over $(0, c_{\max}]$, $a$ ranges over $[a_{\min}, 1)$, where
$a_{\min} = 1-k\cdot c_{\max} = \max\{0, 1-k/X^2\}$. Note that $a_{\min} \ge 0$.

We aim to show that the derivative $R'(c;k,X) \le 0$.

\paragraph{Step 1: Edge Cases ($X=1$ or $k=0$).}

If $X=1$, then $b=0$. $H(c)=k+1-kc$. $\mean_1/c = k+1-kc$.
\[
R(c;k,1) = \frac{(k+1-kc)\left(1-\frac{1}{k+1-kc}\right)}{k(1-c)} = \frac{k+1-kc-1}{k(1-c)} = \frac{k(1-c)}{k(1-c)} = 1.
\]
If $k=0$. $H(c)=X$. $\mean_1/c=X^2$. $b=X(X-1)$.
\[
R(c;0,X) = \frac{X^2(1-1/X)}{b} = \frac{X(X-1)}{X(X-1)} = 1.
\]
In these cases, $R(c;k,X)$ is constant, hence non-increasing.

\paragraph{Step 2: General Case ($X>1, k>0$).}

Through direct differentiation and algebraic manipulation, the derivative is found to be:
\[
R'(c;k,X)
=-\,\frac{k\,(X-1)}{\,H(c)^2\,D(c)^2}\;\Psi(a),
\]
where $D(c):=k(1-c)+(1-kc)X(X-1)$, $a=1-kc$, and $\Psi(a)$ is a quadratic function:
\[
\Psi(a):=A a^2 + B a + C,
\]
with coefficients:
\begin{align*}
A &= X^2(X^2\!-\!X\!+\!k), \\
B &= 2Xk(k\!-\!1), \\
C &= k(k\!-\!1)(k\!-\!X).
\end{align*}

Since $k>0$ and $X>1$, the prefactor $-\frac{k\,(X-1)}{\,H(c)^2\,D(c)^2}$ is negative. To prove $R'(c;k,X) \le 0$, we will show that $\Psi(a) \ge 0$ for all $a \in [a_{\min}, 1)$.

\paragraph{Step 3: Analysis of $\Psi(a)$.}

The leading coefficient is $A = X^2(X(X-1)+k)$. Since $X>1$ and $k>0$, $A>0$. Thus, $\Psi(a)$ is a strictly convex function.

To determine the minimum of a convex function on an interval, we analyze the location of its global minimum (the vertex) $a_v = -B/(2A)$:
\[
a_v = -\frac{2Xk(k-1)}{2X^2(X^2-X+k)} = -\frac{k(k-1)}{X(X^2-X+k)}.
\]
We compare the vertex $a_v$ with the left boundary $a_{\min}$.

\textit{Case 1: $k \ge 1$.}
If $k=1$, $a_v=0$. If $k>1$, $k-1>0$. Since the denominator is positive, $a_v < 0$.
In both scenarios, $a_v \le 0$. Since $a_{\min} \ge 0$, we have $a_v \le a_{\min}$.

\textit{Case 2: $0 < k < 1$.}
Here $k-1<0$, so $a_v > 0$. Since $X>1$, $X^2>1$. Thus $k < X^2$, which means $a_{\min} = 1-k/X^2 > 0$.
We verify the inequality $a_v \le a_{\min}$:
\[
\frac{k(1-k)}{X(X^2-X+k)} \le 1-\frac{k}{X^2} = \frac{X^2-k}{X^2}.
\]
Since all terms are positive, we can cross-multiply and divide by $X$:
\[
X k(1-k) \le (X^2-k)(X^2-X+k).
\]
We analyze the difference RHS - LHS.
\begin{align*}
\text{RHS} &= X^2(X^2-X+k) - k(X^2-X+k) \\
&= X^4-X^3+kX^2 -kX^2+kX-k^2 = X^4-X^3+kX-k^2. \\
\text{LHS} &= Xk - Xk^2. \\
\end{align*}
Therefore,
\begin{align*}
\text{RHS} - \text{LHS} &= (X^4-X^3+kX-k^2) - (Xk-Xk^2) \\
&= X^4-X^3-k^2+Xk^2 \\
&= X^3(X-1)+k^2(X-1) \\
&= (X^3+k^2)(X-1).
\end{align*}
Since $X>1$ and $k>0$, this difference is strictly positive. Thus, $a_v < a_{\min}$.

Therefore, in all cases ($k>0$), the vertex $a_v$ lies to the left of or at the start of the interval $[a_{\min}, 1)$. Since $\Psi(a)$ is convex, it is non-decreasing on this interval. Therefore, the minimum of $\Psi(a)$ on $[a_{\min}, 1)$ occurs at $a_{\min}$.

\paragraph{Step 4: Verification of $\Psi(a_{\min}) \ge 0$.}

It suffices to show that $\Psi(a_{\min}) \ge 0$.

\textit{Case A: $k \ge X^2$ (implies $a_{\min}=0$).}
$\Psi(0) = C = k(k-1)(k-X)$.
Since $X>1$, $X^2>1$. $k\ge X^2$ implies $k>1$, so $k-1>0$.
$k\ge X^2$ and $X\ge 1$ implies $k\ge X$, so $k-X\ge 0$.
Thus, $\Psi(0) \ge 0$.

\textit{Case B: $k < X^2$ (implies $a_{\min}=1-k/X^2$).}
Substituting $a_{\min}$ into $\Psi(a)$ and simplifying yields the identity:
\[
\Psi(a_{\min})=\frac{X-1}{X^2}\cdot P(X,k),
\]
where $P(X,k) = X^5-X^3k+X^2k^2+Xk^3-Xk^2-k^3$.
We must show $P(X,k) \ge 0$. We factor $P(X,k)$ by regrouping terms:
\begin{align*}
P(X,k) &= (X^5-X^3k) + (X^2k^2-Xk^2) + (Xk^3-k^3) \\
&= X^3(X^2-k) + k^2X(X-1) + k^3(X-1) \\
&= X^3(X^2-k) + (X-1)(k^2X+k^3).
\end{align*}
Since we are in Case B, $k < X^2$, so $(X^2-k) > 0$.
Since $X > 1$, $(X-1) > 0$.
Since $k>0$, all terms in the expression are strictly positive.
Therefore, $P(X,k) > 0$, which implies $\Psi(a_{\min}) > 0$.

\paragraph{Step 5: Conclusion.}

We have shown that $\Psi(a) \ge 0$ for all $a \in [a_{\min}, 1)$. Consequently, $R'(c;k,X) \le 0$ for all feasible $c$. The function $R(c;k,X)$ is non-increasing in $c$.
\end{proof}

%% Section: Multiple Actions
\section{Multiple Actions}
\label{apx:multi-action}

In this section, we focus on single-agent environments and consider the extension to multiple actions. 
We show that the loss from full transparency is linear in the number of actions in the worst case, and that this is tight. 
Intuitively, when there are multiple actions, the state can be informative about how the agent could shirk while retaining high utility. 
In this case, providing no information to the agent prevents such shirking incentives, which is why opaqueness can significantly help the principal to obtain a high payoff in comparison to full information.

Consider a version with states of the single-agent multi-action setting (see \cite{dutting2019simple} for the stateless version): In this context, $n$ is the number of actions which the agent has access to. 
Each action $i\in[n]$ is associated with a cost $c_i\geq 0$.
Subtracting $\min_{i\in[n]}c_i$ from every action cost leaves the agent's action choices and the principal's payoff unchanged, so we normalize $\min_{i\in[n]}c_i=0$.
If the agent chooses action $i$, the agent needs to pay a cost of $c_i\geq 0$, and the reward $r$ is drawn according to distribution $F_{\omega,i}$.

\begin{theorem}\label{thm:multi_action}
For single-agent multi-action contract settings with states, it holds that full revelation is a $\Theta(n)$-approximation to the optimal information structure, where $n$ is the number of actions. 
\end{theorem}

\begin{proof}
We first prove the upper bound. 
Let 
\begin{align*}
\Wel_{\omega} = \max_{i\in [n]} \{ \expect[r\sim F_{\omega,i}]{r} - c_i \}
\end{align*}
be the welfare given realized state $\omega$. 
\citep{dutting2019simple} (Theorem 2) show that in the single-agent environment given any state $\omega$, 
there exists a corresponding linear contract $\alpha_{\omega}$ such that 
the principal's payoff given $\alpha_{\omega}$, denoted by $U_P(\alpha_{\omega}, \omega)$, 
is an $O(n)$-approximation to the welfare:
\begin{align*}
O(n) \cdot U_P(\alpha_{\omega}, \omega) \geq \Wel_{\omega}, 
\qquad \forall\omega\in\Omega. 
\end{align*}
By taking the expectation over $\omega$ and observing that the optimal principal's payoff at each state is weakly higher than her payoff under a linear contract, we have
\begin{align*}
O(n) \cdot U^F_P(\prior)\geq
O(n) \cdot \expect[\omega\sim\prior]{U_P(\alpha_{\omega}, \omega)} 
\geq \expect[\omega\sim\prior]{\Wel_{\omega}} 
\geq \OPT(\prior),
\end{align*}
where the last inequality holds since welfare is an upper bound on optimal profit.

Next, we prove the lower bound. Specifically, we construct a family of instances similar to a stateless ``equal revenue'' construction of \citep{DuttingRT21}. Their construction is depicted in Table~\ref{tab:equal-rev}, with $n$ actions and two outcomes, parameterized by $\gamma \in (0,1), \gamma\to 0$: 

\begin{table}[h]
    \centering
    \begin{tabular}{|l||c|c||c|}
    \hline 
    \rule{0pt}{2ex}  \hspace{1.0mm} \hspace{1.0mm}  & \hspace{3.0mm} \bf reward $r_1 = 0$ \hspace{3.0mm} & \hspace{3.0mm} \bf reward $r_2 = \frac{1}{\gamma^{n-1}}$ \hspace{3.0mm} & \hspace{3.0mm} \bf cost \hspace{3.0mm} \\ \hline\hline 
    \rule{0pt}{2ex} \bf action $i \in [n]$: & $F_i(r_1)=(1-\gamma^{n-i}$) & $F_i(r_2)=\gamma^{n-i}$ & $c_i = \frac{1}{\gamma^{i-1}}-i+(i-1)\gamma$ \\ 
    \hline
    \end{tabular}
    \caption{Starting point for lower-bounding the approximation ratio of full transparency.}
    \label{tab:equal-rev}
\end{table}

We adjust the above setting as follows: We create a duplicate $i'$ for each action $i<n$. There are two states with equal prior probabilities, and for each action except for action $n$, one of the duplicates has zero probability for the nonzero reward $r_2$ in each of the states. In particular, in state 1, the probabilities are $F_{1,i}(r_2)=\gamma^{n-i}$ and $F_{1,i'}(r_2)=0$, and in state 2, vice versa --- that is $F_{2,i}(r_2)=0$ and $F_{2,i'}(r_2)=\gamma^{n-i}$.
By revealing full information, we are back in (an equivalent setting to) the setting of \citep{DuttingRT21}, who show that the optimal revenue is~$1$ for the constructed instance, despite $\Wel_\omega=\Omega(n)$.  However, for $n\ge2$ and $0<\gamma\le1/(4n)$, revealing no information and offering $\alpha=1-\frac n2\gamma^{n-1}$ induces action $n$ and gives the principal revenue $n/2$. Indeed, action $n$ gives the agent utility $n/2-(n-1)\gamma>1/2$, every zero-cost action gives at most $1/2$, and every other action $i<n$ gives at most $-\gamma^{-(i-1)}/2+i<0$. The construction has $2n-1$ actions, so the lower bound is linear in the total number of actions.
\end{proof}

\section{An Algorithm for Submodular Team Contracts}
\label{app:algorithms}

We prove \Cref{thm:computation} using value queries. \Cref{app:algorithm-value} presents the algorithm: enumerate agents with large payment requirements, select the remaining agents by greedy additions and pruning, and combine the two groups. \Cref{app:algorithm-analysis} gives its analysis.

Throughout, $\reward$ is nonnegative, monotone, and submodular. As in Section~\ref{sec:multi-agents}, $\kappa(S)$ is the minimum total incentive share and $\revenue(S)=(1-\kappa(S))\reward(S)$ is the principal's utility. Write
\[
w_i=\sqrt{\cost_i},\qquad W(S)=\sum_{i\in S}w_i,
\]
and recall that $\gamma_\star=1-\varpi$, where $\varpi e^{\varpi}=1$. Agents outside the chosen team receive zero shares; zero-cost agents also receive zero shares, including when their marginals vanish. We always retain the empty team as a candidate, evaluate candidate teams at their minimum payments, and use a fixed order for processing agents and breaking ties. The contracts used for our guarantees have total share at most one.

We count value queries and arithmetic operations. For bit complexity, costs, accuracy parameters, and value-oracle answers are rational, and running time is measured in their encoding lengths. A smaller rational accuracy within a factor of two suffices for a real target accuracy. We retain the original baseline $\reward(\emptyset)$. Simultaneously rescaling rewards and costs preserves incentive shares and scales utility by the same factor; hence the pruning bounds also apply to rescaled rewards exceeding one.

\subsection{The algorithm}
\label{app:algorithm-value}

Fix a target loss $\tau\in(0,1/10]$ in approximation fraction. The algorithm below obtains utility at least $(\gamma_\star-\tau)\OPT(\reward,\cost)$ using $n^{\mathrm{poly}(1/\tau)}$ value queries and arithmetic operations. Its small-agent routine uses tolerance $\tau/2$. Let $\eta=\eta(\tau/2)\le1$ be the individual-weight bound in \Cref{lem:alg-diffuse},\footnote{The construction in \Cref{lem:alg-diffuse} uses $\eta=\frac1{32}(\tau^2/1600000)^3$.} and set $\delta=\eta^2/12$. Use a multiplicative grid of cost multipliers $\theta\in[1,3]$, starting at $1$ with ratio $1+\tau/12$.
First discard every positive-cost agent with $\reward(\{i\})=\reward(\emptyset)$: submodularity makes its marginal zero in every team. Among the remaining agents, call a positive-cost agent \emph{large} if
\[
\frac{\cost_i}{\reward(\{i\})-\reward(\emptyset)}>\delta,
\]
and call every other agent \emph{small}. In particular, every zero-cost agent is small. These classes use the original costs and reward function throughout the algorithm.

\begin{enumerate}
\item \emph{Initialize.} Evaluate the empty team and every feasible singleton, retain these candidates, and let $M$ be their largest utility. If $M=0$, return the empty team.

\item \emph{Enumerate the large agents.} For every set $H$ of large agents with $|H|\le\lceil1/\delta\rceil$, retain its minimum-payment contract if feasible, and perform the next two steps for every multiplier $\theta$ in the grid.

\item \emph{Select the small agents at increased costs.} Apply the greedy-and-prune routine described below to the residual reward
\[
T\longmapsto\reward(H\cup T)-\reward(H)
\]
over the small agents, with costs $\theta\cost_i$, tolerance $\tau/2$, and reward-scale interval $[M/2,3nM]$. Denote the returned team by $T_\theta$ and recompute its minimum shares for these increased costs. If their sum exceeds one, skip this candidate. Otherwise, divide the shares by $\theta$ to obtain shares at the original costs, and keep these shares fixed for the next step.

\item \emph{Prune the large agents.} For the selected team $T_\theta$, let $B_0$ be the share left after its fixed payments, and set $V_0=B_0\reward(H\cup T_\theta)$. If $V_0=0$, delete all of $H$. Otherwise, repeatedly delete an agent $i\in H$ whose current marginal in the scaled reward $B_0\reward$ is less than $\sqrt{V_0\cost_i}$, keeping both $B_0$ and $V_0$ fixed throughout these deletions. Pay each remaining large agent its minimum share in the resulting team. If this contract is infeasible or has negative utility, use $T_\theta$ alone. Retain the resulting candidate.

\item \emph{Return the best candidate.} Recompute minimum payments at the original costs and return a feasible candidate of greatest utility.
\end{enumerate}

\paragraph{The greedy-and-prune routine.}
The routine uses weights equal to the square roots of its input costs. It includes all free agents at the start of each run and keeps them fixed. It guesses a reward scale $\widehat F$ in the supplied interval and a parameter $z\in[1/4,5/6]$, skipping guesses that fail the individual-weight cutoff specified below. After dividing rewards and costs by $\widehat F$, it fills a team by repeatedly adding the available agent of highest marginal reward per unit weight, stopping before the next addition would exceed the fixed weight cap $-z\log z$. It then processes the positive-cost agents in that team in a fixed order, retaining or deleting each one and recording the share needed when an agent is retained. The decisions use a quadratic potential that controls payment, retained weight, and reward loss together. If the pruned reward reaches the run's target, the run stops; otherwise, density greedy refills the pruned team to the same cap, allowing deleted agents to return, and pruning repeats. Each run has a prescribed polynomial bound on its number of rounds, and all resulting feasible contracts are retained.

The routine also makes one unrestricted density-greedy pass and considers every prefix, including the initial free team. It applies the original Potential Lemma with benchmark shares proportional to the weights, subject to the numerical safeguards in \Cref{apx:alg-diffuse}. It returns the best candidate from these prefixes, the refill runs, and the free team. \Cref{apx:alg-diffuse} gives the finite parameter grids, stopping target, potential, and numerical precision, and proves the guarantee in \Cref{lem:alg-diffuse}.

\subsection{Analysis}
\label{app:algorithm-analysis}

We prove the utility guarantee $(\gamma_\star-\tau)\OPT(\reward,\cost)$ and then derive the two approximation factors in \Cref{thm:computation}.

\subsubsection{Why the enumeration and reward scales suffice}
\label{app:algorithm-preliminaries}

We first show that the enumeration includes the large agents of an optimal team and establish the reward and cost bounds needed for the small-agent routine.

\paragraph{A computable lower bound on the optimum.}
The initialization step computes $M$ from the empty team and feasible singletons. This value bounds the optimum as follows:
\begin{equation}\label{eq:alg-anchor}
M\le\OPT(\reward,\cost)\le nM.
\end{equation}
To prove the upper bound, fix a nonempty optimal team $O$ with positive utility. Let
\[
k_i=\frac{\cost_i}{\reward(i\mid O\setminus\{i\})}
\]
be agent $i$'s share in this team, with $k_i=0$ for zero-cost agents. Using the notation from Section~\ref{sec:multi-agents}, its total share is $\kappa(O)=\sum_{i\in O}k_i<1$, and
\[
\OPT(\reward,\cost)=(1-\kappa(O))\reward(O).
\]
Removing the other agents weakly increases $i$'s marginal, so its singleton utility is at least $(1-k_i)\reward(\{i\})$. Since $\sum_{i\in O}\reward(\{i\})\ge\reward(O)$ and $\reward(\{i\})\le\reward(O)$,
\[
|O|M\ge\sum_{i\in O}(1-k_i)\reward(\{i\})
\ge(1-\kappa(O))\reward(O).
\]
The empty optimal team is already a candidate. Hence $M=0$ makes the problem trivial. For the remaining analysis, we assume $M>0$ and that the empty candidate is not optimal, so $O$ can be taken nonempty with positive utility.

\paragraph{Enumerating agents with large payment requirements.}
By the classification above, every large agent in $O$ has $k_i>\delta$, so fewer than $1/\delta$ large agents belong to $O$. Enumerating all large-agent sets of size at most $\lceil1/\delta\rceil$ therefore includes
\[
H=\{i\in O:i\text{ is large}\},\qquad L=O\setminus H.
\]
For this enumerated set, Cauchy--Schwarz and submodularity give
\begin{equation}\label{eq:alg-root-bounds}
\begin{aligned}
W(H)^2&\le\reward(H)\sum_{i\in H}k_i,\\
W(L)^2&\le\bigl(\reward(O)-\reward(H)\bigr)\sum_{i\in L}k_i.
\end{aligned}
\end{equation}
Indeed, $w_i^2=k_i\reward(i\mid O\setminus\{i\})$, while the optimal marginals sum to at most $\reward(H)$ over $H$, and to at most $\reward(O)-\reward(H)$ over $L$. Throughout the analysis, $k_i$ remains the share required in $O$, even when we consider a smaller team.

This classification also gives a bound on every small agent's cost. Its singleton contract pays a share at most $\delta$, and hence has utility at least
$(1-\delta)(\reward(\{i\})-\reward(\emptyset))$. It follows that
\begin{equation}\label{eq:alg-small-cost}
\cost_i\le\frac{\delta M}{1-\delta}
\qquad\text{for every small agent }i.
\end{equation}
These definitions use the original reward function, including its baseline $\reward(\emptyset)$. Zero-cost agents are small, but we do not require the optimal team to contain all of them.

\paragraph{When a singleton already suffices.}
The singleton candidates let us restrict attention to optimal contracts that leave the principal at least a third of the reward, which bounds $\reward(O)$ in terms of $M$.

\begin{lemma}\label{lem:alg-large-payment}
If $\kappa(O)\ge2/3$, some singleton has utility at least $\kappa(O)\OPT(\reward,\cost)$. Consequently, unless the singleton candidates already give a $2/3$-fraction of the optimum, we may assume
\begin{equation}\label{eq:alg-reward-scale}
\kappa(O)<\frac23,\qquad \reward(O)\le3\OPT(\reward,\cost)\le3nM.
\end{equation}
\end{lemma}
\begin{proof}
Deleting agent $i$ from $O$ preserves the other agents' incentives at their previous shares. Optimality therefore implies
\[
(1-\kappa(O))\reward(O)
\ge(1-\kappa(O)+k_i)\reward(O\setminus\{i\}),
\]
or, equivalently,
\[
\frac{\reward(i\mid O\setminus\{i\})}{\reward(O)}
\ge\frac{k_i}{1-\kappa(O)+k_i}.
\]
Choose $i^\star\in O$ with the largest share $k_{i^\star}$. Summing the last inequality and using submodularity gives
\[
1\ge\sum_{i\in O}\frac{k_i}{1-\kappa(O)+k_i}
\ge\frac{\kappa(O)}{1-\kappa(O)+k_{i^\star}},
\]
so $k_{i^\star}\ge2\kappa(O)-1$. The singleton $\{i^\star\}$ has utility at least
\[
\frac{k_{i^\star}(1-k_{i^\star})}{1-\kappa(O)+k_{i^\star}}\reward(O).
\]
For $\kappa(O)\ge2/3$, we have
$(1-\kappa(O))^2\le2\kappa(O)-1\le k_{i^\star}\le\kappa(O)$. Consequently,
\begin{align*}
&k_{i^\star}(1-k_{i^\star})
-\kappa(O)(1-\kappa(O))(1-\kappa(O)+k_{i^\star})\\
&\qquad=(\kappa(O)-k_{i^\star})
\bigl(k_{i^\star}-(1-\kappa(O))^2\bigr)\ge0.
\end{align*}
The displayed singleton utility is therefore at least $\kappa(O)\OPT(\reward,\cost)$.
\end{proof}

\subsubsection{Guarantee of the greedy-and-prune routine}
\label{apx:alg-diffuse}

We analyze the routine in Step~3 against a comparison team $P$ satisfying $W(P)^2\le k\reward(P)$. Its guarantee is a fraction of $(1-k)\reward(P)$, allowing us to apply it at increased costs. The bound requires small individual weights, which motivates the initial classification of agents.

\begin{lemma}[The small-agent algorithm]
\label{lem:alg-diffuse}
For every $\zeta\in(0,1/10]$, there is $\eta=\Theta(\zeta^6)>0$ and a deterministic value-oracle algorithm with the following guarantee. Given $0<F_{\min}\le F_{\max}$, it returns a feasible contract of utility at least
\begin{equation}\label{eq:alg-diffuse-guarantee}
 (\gamma_\star-\zeta)(1-k)\reward(P)
\end{equation}
for every comparison team $P$ and $k\in[0,1]$ satisfying
\[
 \reward(P)\in[F_{\min},F_{\max}],\qquad
 W(P)^2\le k\reward(P),\qquad \max_iw_i\le\eta\sqrt{\reward(P)}.
\]
Its query and arithmetic-operation counts are polynomial in $n$, $1/\zeta$, and $1+\log(F_{\max}/F_{\min})$.
\end{lemma}

Density greedy supplies a reward-growth estimate for both candidate families. The original Potential Lemma controls the utility of greedy prefixes, while a second pruning bound also controls the weight retained between refill rounds.

\paragraph{Greedy growth.}
Let $S_0,S_1,\ldots$ be the teams in a greedy run, where $S_0$ contains all free agents. At step $j+1$, add an available positive-cost agent of maximum density $\reward(i\mid S_j)/w_i$ to obtain $S_{j+1}$. Free agents remain fixed, and their contribution is incorporated into the reward function passed to the pruning routines.

Suppose that $z>0$, $\reward(P)\ge\widehat F>0$ and $W(P)\le z\sqrt{\widehat F}$. Whenever $\reward(S_j)<\widehat F$, submodularity gives
\[
 \sum_{i\in P\setminus S_j}\reward(i\mid S_j)
 \ge\widehat F-\reward(S_j).
\]
Thus the next agent has density at least $(\widehat F-\reward(S_j))/(z\sqrt{\widehat F})$. Multiplying the residual bounds over the additions gives, for any two steps $a\le b$,
\begin{equation}\label{eq:alg-greedy-growth}
 \reward(S_b)\ge\widehat F-(\widehat F-\reward(S_a))_+
 \exp\!\left(-\frac{W(S_b)-W(S_a)}{z\sqrt{\widehat F}}\right).
\end{equation}
If an added agent has weight exceeding $z\sqrt{\widehat F}$, that addition already raises the reward to at least $\widehat F$. If all agents have been added, the reward is also at least $\widehat F$.

In particular, consider a run starting from the free agents alone, so $W(S_0)=0$. If $w_i\le\eta\sqrt{\widehat F}$ for every agent, then for any $t>0$ the first prefix crossing weight $tz\sqrt{\widehat F}$ has reward at least $(1-e^{-t})\widehat F$ and weight at most $(tz+\eta)\sqrt{\widehat F}$. If there is no such prefix, the entire pool has reward at least $\widehat F$ and smaller weight. Including free agents does not affect this comparison, since it can only increase the reward of the comparison team.

\paragraph{Pruning the greedy prefixes.}
The routine evaluates each greedy prefix by pruning it into a feasible contract. The following form of \Cref{lem:pruning} expresses the resulting utility directly in terms of the prefix's reward and weight.

\begin{lemma}[Square-root form of the Potential Lemma]
\label{lem:alg-square-pruning}
Given a team $S$, the pruning procedure in exact real arithmetic uses $O(n^2)$ value queries and finds a feasible contract with utility at least
\begin{equation}\label{eq:alg-square-pruning}
 \left(\sqrt{\reward(S)}-\frac{W(S)}2\right)_+^2-\sum_{i\in S}\cost_i.
\end{equation}
Zero-cost agents in $S$ may be kept fixed.
\end{lemma}
\begin{proof}
Incorporate those zero-cost agents into the baseline for this run. The claim is immediate if $\reward(S)=0$ or $W(S)>2\sqrt{\reward(S)}$, since its lower bound is nonpositive. Otherwise apply \Cref{lem:pruning} with benchmark shares $\kappa_i=w_i/(2\sqrt{\reward(S)})$ for the positive-cost agents. Its principal term is the square in \eqref{eq:alg-square-pruning}, and its correction term is at least $-\sum_{i\in S}\cost_i$, because all marginal contributions are nonnegative.
\end{proof}

The exact bound is used in the analysis; the polynomial-precision implementation for the prefix candidates is given in the numerical-precision discussion below.

\paragraph{Pruning while controlling the retained weight.}
The prefix candidates suffice when $k$ is near zero or one. For the remaining values, the refill runs need to control the retained weight together with reward and payments. The next lemma provides this control after normalizing the reward scale. Its proof uses the same retain-or-delete averaging idea as \Cref{lem:pruning}, with a quadratic potential guiding the deterministic choices.

\begin{lemma}[Simultaneous pruning]
\label{lem:alg-simultaneous-pruning}
Let $S$ consist of positive-cost agents, with $0<W(S)\le1$. Fix $\xi\in(0,1)$ and suppose
\begin{equation}\label{eq:alg-pruning-smallness}
 (2+\xi^{-1})\sum_{i\in S}w_i^2\le\frac{\xi^2}{2}.
\end{equation}
A deterministic deletion algorithm returns $T\subseteq S$ and recorded shares supporting its agents such that
\begin{align}
 \sum_{i\in T}\contract_i&\le W(S)+\xi,
 \label{eq:alg-pruning-share}\\
 \reward(T)&\ge\reward(S)-W(T)(1-W(T)/W(S))-3\xi.
 \label{eq:alg-pruning-reward}
\end{align}
The algorithm uses $O(n)$ value queries. Each comparison of its quadratic potential may have additive error at most $\xi^2/(2n)$.
\end{lemma}
\begin{proof}
Fix an ordering $i_1,\ldots,i_{|S|}$ of the agents, and let $S_0=S$. At step $j$, compute the processing marginal
\[
 m_j=\reward(i_j\mid S_{j-1}\setminus\{i_j\}),
 \qquad r_j=\min\{1,m_j/w_{i_j}\},
\]
then replace $r_j$ by zero if $r_j<\xi$. Choose $X_j\in\{0,1\}$, taking $X_j=r_j$ when $r_j\in\{0,1\}$ and using the potential comparison below otherwise. Set
\[
 S_j=\begin{cases}
 S_{j-1},&X_j=1,\\
 S_{j-1}\setminus\{i_j\},&X_j=0.
 \end{cases}
\]
When $X_j=1$, record $\contract_{i_j}=\cost_{i_j}/m_j$; set $\contract_{i_j}=0$ when $X_j=0$. The quantities $m_j$, $r_j$, and the recorded share are snapshots from step $j$ and are not recomputed at later steps. The output team is $T=S_{|S|}$.

Initialize $E_{w,0}=E_{f,0}=E_{p,0}=\Phi_0=0$. After step $j$, update the centered errors in retained weight, reward loss, and payment by
\begin{align*}
 E_{w,j}&=E_{w,j-1}+w_{i_j}(X_j-r_j),\\
 E_{f,j}&=E_{f,j-1}-m_j(X_j-r_j),\\
 E_{p,j}&=E_{p,j-1}+\frac{\cost_{i_j}}{m_j}(X_j-r_j),\\
 \Phi_j&=E_{w,j}^2+E_{f,j}^2+E_{p,j}^2.
\end{align*}
The payment increment is defined to be zero when $r_j=0$; below, the averaged payment $r_j\cost_{i_j}/m_j$ is also defined to be zero in this case. Thus these expressions never require division by a zero marginal. At a fractional choice, let $\Phi_j^{(x)}$ be the potential obtained from these updates with $X_j=x$, keeping the state after step $j-1$ fixed. Choose $X_j$ so that
\[
 \Phi_j\le\min\{\Phi_j^{(0)},\Phi_j^{(1)}\}
                  +\frac{\xi^2}{2n}.
\]

To bound this choice, average the two candidate updates with weights $r_j$ and $1-r_j$. Conditional on the decisions through step $j-1$, each centered increment has mean zero, so its cross term with the preceding error cancels. At a fractional choice, $m_j=w_{i_j}r_j$ and $r_j\ge\xi$, whence
\begin{align*}
 r_j\Phi_j^{(1)}+(1-r_j)\Phi_j^{(0)}
 &=\Phi_{j-1}+r_j(1-r_j)
   \left(w_{i_j}^2+m_j^2+\frac{w_{i_j}^4}{m_j^2}\right)\\
 &\le\Phi_{j-1}+(2+\xi^{-1})w_{i_j}^2.
\end{align*}
For $r_j\in\{0,1\}$, all three centered increments vanish. Therefore, at every step,
\[
 \Phi_j\le\Phi_{j-1}+(2+\xi^{-1})w_{i_j}^2
                         +\frac{\xi^2}{2n}.
\]
Summing along the actual sequence of decisions and using $|S|\le n$ and \eqref{eq:alg-pruning-smallness} gives
\[
 \Phi_{|S|}\le(2+\xi^{-1})\sum_{i\in S}w_i^2
                  +\frac{|S|\xi^2}{2n}\le\xi^2.
\]
Consequently, each final centered error has absolute value at most $\xi$, even though $m_j$ and $r_j$ depend on the preceding decisions.

By the weight-error updates,
\[
 W(T)=\sum_{j=1}^{|S|}w_{i_j}r_j+E_{w,|S|}.
\]
Each averaged payment is at most $w_{i_j}$: it is zero when $r_j=0$, equals $w_{i_j}$ when $0<r_j<1$, and is at most $w_{i_j}$ when $r_j=1$. The payment-error updates therefore give
\[
 \sum_{i\in T}\contract_i
 =\sum_{j=1}^{|S|}r_j\frac{\cost_{i_j}}{m_j}+E_{p,|S|}
 \le W(S)+\xi,
\]
which proves \eqref{eq:alg-pruning-share}.

The reward-error updates and the successive deletions give the exact identity
\[
 \reward(S)-\reward(T)
 =\sum_{j=1}^{|S|}m_j(1-X_j)
 =\sum_{j=1}^{|S|}m_j(1-r_j)+E_{f,|S|}.
\]
If $r_j=0$, then $m_j<\xi w_{i_j}$; if $0<r_j<1$, then $m_j=w_{i_j}r_j$; and if $r_j=1$, the averaged loss is zero. Thus weighted Jensen gives
\begin{align*}
 \sum_{j=1}^{|S|}m_j(1-r_j)
 &\le\sum_{j=1}^{|S|}w_{i_j}r_j(1-r_j)+\xi W(S)\\
 &\le\left(\sum_{j=1}^{|S|}w_{i_j}r_j\right)
 \left(1-\frac{\sum_{j=1}^{|S|}w_{i_j}r_j}{W(S)}\right)+\xi W(S).
\end{align*}
Both $W(T)$ and $\sum_{j=1}^{|S|}w_{i_j}r_j$ lie in $[0,W(S)]$, where $x\mapsto x(1-x/W(S))$ is $1$-Lipschitz. Replacing the latter sum by $W(T)$ therefore costs at most $|E_{w,|S|}|\le\xi$. Adding $|E_{f,|S|}|\le\xi$ and using $W(S)\le1$ proves \eqref{eq:alg-pruning-reward}.

Finally, $T\setminus\{i_j\}\subseteq S_{j-1}\setminus\{i_j\}$ for every retained agent $i_j$, so submodularity gives $\reward(i_j\mid T\setminus\{i_j\})\ge m_j>0$. Hence its recorded share supports effort in the final team. The smallness condition also gives $w_{i_j}<\xi$, so every retained share is at most $w_{i_j}/\xi<1$. Storing the reward of the current team, each step needs just one further value query to obtain $m_j$ and the reward after a possible deletion. Including the initial query, this uses $|S|+1=O(n)$ value queries.
\end{proof}

\begin{proof}[Proof of \Cref{lem:alg-diffuse}]
The algorithm keeps the best feasible minimum-payment contract found in the procedures below, together with the zero-payment contract under which all free agents work.

\paragraph{The refill argument.}
Suppose first that a reward lower bound $\widehat F$ and a parameter $z\in[1/4,5/6]$ are supplied, with $\reward(P)\ge\widehat F$ and $W(P)\le z\sqrt{\widehat F}$. Divide rewards and costs by $\widehat F$. In these units, the comparison reward is at least one and its weight is at most $z$.
Choose the weight cap
\begin{equation}\label{eq:alg-refill-cap}
 C_z=-z\log z.
\end{equation}
The cap $C_z$ is fixed throughout this run. It makes the greedy reward bound $1-e^{-C_z/z}=1-z$. Pruning a team of this weight pays at most approximately $C_z$. Thus a retained reward close to $1-z$ gives utility close to $(1-C_z)(1-z)$.
When pruning deletes substantial weight, a greedy refill recovers more reward than pruning loses. We make this progress argument precise below, keeping track of the team before and after each round.

Set
\[
 \nu=\zeta/20,\qquad \xi=\nu^2/1000,
 \qquad \eta=\xi^3/32.
\]
For the supplied-parameter construction, assume every normalized weight is at most $4\eta$. Index refill rounds by $t=0,1,\ldots$. Let $S_t$ be the team before pruning in round $t$, and let $T_t$ be the team it retains. Construct $S_0$ from the free agents by density greedy, stopping before the next addition would exceed the cap $C_z$. Allow a conservative implementation that leaves a deficit of at most $8\eta$ unless the pool is exhausted. By \eqref{eq:alg-greedy-growth}, the resulting team satisfies
\begin{equation}\label{eq:alg-initial-reward}
 \reward(S_0)\ge1-z-32\eta,
\end{equation}
since the growth bound changes by at most $8\eta/z\le32\eta$ under this deficit. An exhausted pool instead has reward at least one.

In round $t$, if $W(S_t)=0$, set $T_t=S_t$ and assign zero shares. Otherwise, apply \Cref{lem:alg-simultaneous-pruning} to $S_t$, folding its fixed free agents into the baseline. Its smallness condition holds because $W(S_t)\le C_z\le1$ and
\[
 (2+\xi^{-1})\sum_{i\in S_t}w_i^2
 \le(2+\xi^{-1})4\eta
 \le3\xi^2/8<\xi^2/2.
\]
Write $D_t=W(S_t\setminus T_t)$ for the weight deleted in round $t$, and $\contract_i^{(t)}$ for the shares recorded in that round. Then
\begin{equation}\label{eq:alg-one-prune}
 \reward(T_t)\ge \reward(S_t)-D_t+D_t^2/C_z-3\xi,\qquad \sum_{i\in T_t}\contract_i^{(t)}\le C_z+\xi.
\end{equation}
For $W(S_t)>0$, the lemma's loss is $D_t-D_t^2/W(S_t)\le D_t-D_t^2/C_z$, since $W(S_t)\le C_z$. When $W(S_t)=0$, the same bounds hold directly with $D_t=0$. The shares are feasible because $C_z\le3/8$ and $\xi<10^{-3}$.
Record this candidate before any further refill. If
\begin{equation}\label{eq:alg-refill-success}
 \reward(T_t)\ge1-z-\nu,
\end{equation}
it has utility at least $(1-C_z-\xi)(1-z-\nu)$. Otherwise, refill from $T_t$ to the same cap to obtain $S_{t+1}$ and repeat, allowing previously deleted agents to be added again. Every run stops after at most $\lceil112/(3\nu^2)\rceil+2$ pruning rounds, keeping all candidates found. This limit applies to every parameter guess, including guesses for which no comparison team satisfies the supplied bounds. We now show that a run with valid bounds succeeds before this limit.

Assume inductively that $\reward(S_t)\ge1-z-32\eta$, as established for $t=0$. A failed round must then delete
\[
 D_t\ge\nu-32\eta-3\xi\ge\nu/2.
\]
The refill adds at least $D_t-8\eta$ weight unless it exhausts the pool. For an exact refill of weight $D_t$, \eqref{eq:alg-greedy-growth} and $1-e^{-x}\ge x-x^2/2$ give reward at least
$\reward(T_t)+(1-\reward(T_t))(D_t/z-D_t^2/(2z^2))$.
The coefficient of $1-\reward(T_t)$ is nonnegative because $D_t/z\le C_z/z\le\log4<7/5$.
Failure implies $1-\reward(T_t)>z+\nu$, and the weight deficit loses at most $32\eta$ reward. If the pool is not exhausted, combining these facts with \eqref{eq:alg-one-prune} yields
\begin{align}
 \reward(S_{t+1})-\reward(S_t)
 &\ge D_t^2\left(\frac1{C_z}-\frac1{2z}\right)-3\xi-32\eta\notag\\
 &\ge\frac3{14}D_t^2-3\xi-32\eta
 \ge\frac{3\nu^2}{112}.
 \label{eq:alg-refill-progress}
\end{align}
Here $1/C_z-1/(2z)\ge3/(14z)\ge3/14$, and the last inequality uses $D_t\ge\nu/2$ and our choices of $\xi$ and $\eta$.
Every failed round also satisfies
\[
 \reward(S_t)<1-z-\nu+C_z/4+3\xi<1,
\]
by \eqref{eq:alg-one-prune}. Thus exhaustion forces the next pruning step to succeed, and there cannot be more than
$\lceil112/(3\nu^2)\rceil+1$ failed rounds. The progress bound preserves $\reward(S_{t+1})\ge1-z-32\eta$, closing the induction.

The successful contract has normalized utility at least
\begin{equation}\label{eq:alg-central-target}
 (1-C_z-\xi)(1-z-\nu)
 \ge(1+z\log z)(1-z)-2\nu
 \ge\gamma_\star(1-z^2)-2\nu.
\end{equation}
To verify the last inequality, differentiate $(1+z\log z)/(1+z)$. The derivative has the sign of $\log z+z$, so the minimum occurs at $z=\varpi$ and equals $1-\varpi=\gamma_\star$.

\paragraph{Guessing the reward and weight scales.}
Run the construction for $\widehat F=F_{\min}(1+\xi)^j\le F_{\max}$, with integers $j\ge0$, and a mesh of spacing at most $\xi$ in $[1/4,5/6]$, including both endpoints. Skip a run if any weight exceeds $4\eta\sqrt{\widehat F}$.
For $k\in[1/16,2/3]$, one downward reward guess and upward mesh point satisfy
\[
 \frac{\reward(P)}{1+\xi}\le\widehat F\le \reward(P),
 \qquad \sqrt{k\reward(P)/\widehat F}\le z\le\sqrt{k\reward(P)/\widehat F}+\xi.
\]
The mesh point lies in $[1/4,5/6]$ because $\xi<10^{-3}$, and this run is not skipped since $w_i/\sqrt{\widehat F}\le\eta\sqrt{1+\xi}<4\eta$.
These choices give $z^2\widehat F\le k\reward(P)+3\xi \reward(P)$, hence
$(1-z^2)\widehat F\ge(1-k)\reward(P)-4\xi \reward(P)$.
Restoring units in \eqref{eq:alg-central-target}, the resulting utility is at least
\[
 \gamma_\star(1-k)\reward(P)-(4\gamma_\star\xi+2\nu)\reward(P)
 \ge(\gamma_\star-\zeta)(1-k)\reward(P),
\]
where the last inequality uses $1-k\ge1/3$. This proves the desired bound on the central range with finite grids.

\paragraph{The endpoint ranges.}
For the remaining values of $k$, run density greedy once on the whole pool and apply \Cref{lem:alg-square-pruning} to every prefix, including the initial free team. For $k>0$ and any $t>0$, consider the first prefix $S$ crossing $t\sqrt{k\reward(P)}$, or the exhausted pool if there is none. Greedy growth and $\sum_{i\in S}\cost_i\le\eta\sqrt{\reward(P)}\,W(S)$ give a pruned candidate $T$ with
\begin{equation}\label{eq:alg-prefix-bound}
 \frac{\revenue(T)}{\reward(P)}\ge
 \left(\sqrt{1-e^{-t}}-\frac{t\sqrt k+\eta}{2}\right)_+^2
 -\eta(t\sqrt k+\eta).
\end{equation}
When $k=0$, the comparison team is free and the zero-payment candidate has utility at least $\reward(P)$.

For $0<k\le1/16$, take $t=6/5$. Since $\sqrt{1-e^{-6/5}}>5/6$, the bound in \eqref{eq:alg-prefix-bound} is at least
$(41/60-\eta/2)^2-\eta(3/10+\eta)>0.45$ for $\eta\le10^{-3}$, exceeding $\gamma_\star(1-k)$ because $\gamma_\star<0.44$.
For $2/3\le k\le1$, take $t=3/5$ and use $\sqrt{1-e^{-3/5}}>2/3$ to obtain
\[
 \frac{\revenue(T)}{\reward(P)}\ge
 \left(\frac23-\frac3{10}\sqrt k\right)^2-2\eta
 \ge\frac12(1-k)+\frac1{225}-2\eta
 \ge\gamma_\star(1-k).
\]
For the middle inequality, subtracting $(1-k)/2$ gives
$59k/100-2\sqrt k/5-1/18$. This expression increases for $k\ge2/3$ and is at least $1/225$ there, using $\sqrt{2/3}\le5/6$.
This fixed additive slack also covers $k$ arbitrarily close to one.

\paragraph{Query count and numerical precision.}
A refill uses $O(n^2)$ value queries and simultaneous pruning uses $O(n)$. There are $O(\nu^{-2})$ rounds per run, $O(1/\xi)$ weight guesses, and $O(1+\log(F_{\max}/F_{\min})/\xi)$ reward guesses. Pruning all greedy prefixes adds $O(n^3)$ queries. As $\eta=\Theta(\zeta^6)$, these counts have the stated polynomial bounds.

To implement a central run in the original weight units, approximate the cap $C_z\sqrt{\widehat F}$ from below with error at most $\eta\sqrt{\widehat F}$, and round the weights upward with total error at most $\eta\sqrt{\widehat F}$ over the ground set. Stop before their rounded sum exceeds the rounded cap. Since each actual weight is at most $4\eta\sqrt{\widehat F}$, the actual deficit is at most $8\eta\sqrt{\widehat F}$, as allowed in the proof. Density and cutoff comparisons reduce to squared comparisons of nonnegative rational quantities. Potential comparisons only need additive precision $\xi^2/(2n)$. No exact comparison of sums of square roots is required.

For the prefix applications of \Cref{lem:alg-square-pruning}, the prefixes used in the proof satisfy $W(S)/\sqrt{\reward(S)}<0.91$. Discard zero-reward prefixes. For every other prefix, compute a rational upper bound on $\sum_{i\in S}\sqrt{\cost_i/\reward(S)}$ with additive error at most $1/100$, by rounding each summand upward with error at most $1/(100n)$. Discard the prefix if this bound exceeds one. This fixed-precision test always terminates, and all prefixes used in the proof pass it. Approximate the ideal benchmark shares $w_i/(2\sqrt{\reward(S)})$ downward within relative error
\[
 \min\left\{\frac1{100},\frac{\xi F_{\min}}{4\reward(S)}\right\}.
\]
The rational shares sum to at most $1/2$. Their benchmark-marginal sum increases by at most the reciprocal of one minus this relative error, while their total share decreases. The square-root guarantee therefore loses at most $\xi F_{\min}$. This is absorbed by the slack in both endpoint ranges: $1/225-2\eta-\xi>0$, and the small-$k$ bound remains above $0.45$. The original pruning algorithm now uses rational quantities throughout. All computations require polynomial numerical precision in the rational input encoding and in $\log(n/\zeta)$.
\end{proof}

\subsubsection{Cost scaling and the final utility guarantee}
\label{app:algorithm-lifting}
\label{app:algorithm-completion}

We now apply the small-agent guarantee in Step~3 and account for the large agents' payments in Step~4. Restoring the original costs divides the returned team's required shares by the cost multiplier, leaving more reward to pay the large agents. Write
\[
b=\gamma_\star-\tau/2\in[1/3,1/2].
\]

\paragraph{Why the small-agent guarantee applies.}
Consider the enumerated set $H$ belonging to an optimal team $O$, with notation as in \Cref{app:algorithm-preliminaries}. The singleton candidates suffice if $\kappa(O)\ge2/3$. Also, the candidate $H$ alone has utility at least
\[
\left(1-\sum_{i\in H}k_i\right)\reward(H),
\]
so it gives the desired fraction $b$ if $\reward(H)\ge b\reward(O)$. We may therefore assume
\[
\kappa(O)<2/3,\qquad \reward(H)<b\reward(O)\le\reward(O)/2.
\]
The residual comparison team $L$ has reward
\[
\reward(H\cup L)-\reward(H)=\reward(O)-\reward(H)\in[M/2,3nM].
\]
By \eqref{eq:alg-small-cost}, every small agent, under every multiplier $\theta\le3$, satisfies
\[
\frac{\theta\cost_i}{\reward(O)-\reward(H)}
\le\frac{6\delta}{1-\delta}\le\eta^2.
\]
Moreover, \eqref{eq:alg-root-bounds} gives the scaled weight bound
\[
\theta W(L)^2\le\theta\left(\sum_{i\in L}k_i\right)
\bigl(\reward(O)-\reward(H)\bigr).
\]
Thus, whenever $\theta\sum_{i\in L}k_i<1$, \Cref{lem:alg-diffuse} returns a small team $T_\theta$ whose minimum scaled shares sum to some $s_\theta<1$ and satisfy
\begin{equation}\label{eq:alg-scaled-guarantee}
(1-s_\theta)\bigl(\reward(H\cup T_\theta)-\reward(H)\bigr)
\ge b\left(1-\theta\sum_{i\in L}k_i\right)
\bigl(\reward(O)-\reward(H)\bigr).
\end{equation}

\paragraph{Choosing the cost multiplier.}
At the original costs, the same team needs total share $s_\theta/\theta$. Its value before paying the large agents is therefore
\begin{equation}\label{eq:alg-virtual-value}
V_\theta=\left(1-\frac{s_\theta}{\theta}\right)\reward(H\cup T_\theta).
\end{equation}
This expression restores the reward $\reward(H)$ that was removed in defining the residual function. Substituting \eqref{eq:alg-scaled-guarantee}, then applying the arithmetic--geometric mean inequality to the terms proportional to $1-s_\theta$ and $(1-s_\theta)^{-1}$, yields
\begin{equation}\label{eq:alg-scaling-bound}
V_\theta\ge
\frac{\left(
\sqrt{b\bigl(\reward(O)-\reward(H)\bigr)
\left(1-\theta\sum_{i\in L}k_i\right)}+
\sqrt{\reward(H)(\theta-1)}\right)^2}{\theta}
=: \mathcal L(\theta).
\end{equation}
The purpose of cost scaling is now explicit: choose $\theta$ to maximize this guaranteed value before paying $H$.

To find the best multiplier, rewrite the last expression as
\[
\mathcal L(\theta)=
\left(
\sqrt{b\bigl(\reward(O)-\reward(H)\bigr)}
\sqrt{\frac1\theta-\sum_{i\in L}k_i}
+\sqrt{\reward(H)}\sqrt{1-\frac1\theta}
\right)^2.
\]
The two quantities under the latter square roots sum to $1-\sum_{i\in L}k_i$. Cauchy--Schwarz therefore bounds this expression by
\[
\left(1-\sum_{i\in L}k_i\right)
\bigl[b\reward(O)+(1-b)\reward(H)\bigr],
\]
with equality at
\begin{equation}\label{eq:alg-ideal-multiplier}
\theta^*=\frac{b\bigl(\reward(O)-\reward(H)\bigr)+\reward(H)}
{b\bigl(\reward(O)-\reward(H)\bigr)+\reward(H)\sum_{i\in L}k_i}.
\end{equation}
This multiplier is in the searched range: $\theta^*\ge1$,
\[
\theta^*\le1+\frac{\reward(H)}{b\bigl(\reward(O)-\reward(H)\bigr)}
\le1+\frac1{1-b}\le3,
\]
and $\theta^*\sum_{i\in L}k_i<1$ since $\sum_{i\in L}k_i<1$. Hence the largest lower bound is
\begin{equation}\label{eq:alg-restored-baseline}
\mathcal L(\theta^*)
=\left(1-\sum_{i\in L}k_i\right)
\bigl[b\reward(O)+(1-b)\reward(H)\bigr].
\end{equation}

\paragraph{Paying the large agents.}
Step~4 converts the value left after paying the small agents into principal utility. Removing large agents preserves the small agents' incentives, so their shares can remain fixed. The following lemma bounds the resulting utility.

\begin{lemma}\label{lem:alg-reserved-pruning}
Let $H$ and $T$ be disjoint teams, with strictly positive costs on $H$. Suppose fixed shares support the agents in $T$ in the team $H\cup T$, and these shares sum to $1-B_0$, where $B_0\in[0,1]$. Put
\[
V_0=B_0\reward(H\cup T),\qquad
\psi_H(v)=\bigl(v-W(H)\sqrt v\bigr)_+.
\]
Both $B_0$ and $V_0$ remain fixed during this call. A polynomial-time deletion procedure returns a subset of $H$ whose union with $T$ can be supported with principal utility at least $\psi_H(V_0)$. Moreover, $\psi_H$ is nondecreasing and $1$-Lipschitz on $[0,\infty)$.
\end{lemma}
\begin{proof}
If $V_0=0$, delete all of $H$. Otherwise, let $H_0=H$ and let $H_j$ be the large agents remaining after $j$ deletions; the small team $T$ stays fixed. If an agent $i\in H_j$ satisfies
\[
B_0\reward\bigl(i\mid(H_j\cup T)\setminus\{i\}\bigr)
<\sqrt{V_0\cost_i},
\]
set $H_{j+1}=H_j\setminus\{i\}$. The marginal changes with $j$, while $B_0$ and the threshold $\sqrt{V_0\cost_i}$ do not. At termination, let $H_J$ be the remaining large agents. Evaluate the contract on $H_J\cup T$, keeping the small agents' shares fixed and using minimum payments for $H_J$. If it is infeasible or has negative utility, use $T$ alone.

The fallback has nonnegative utility, so the conclusion is immediate when $W(H)>\sqrt{V_0}$. Suppose $W(H)\le\sqrt{V_0}$, and consider the terminal set $H_J$. Each deletion loses at most $\sqrt{V_0}\,w_i$ in scaled reward. Each retained agent needs a share at most $w_i/\sqrt{V_0}$ in the scaled problem. Thus the final scaled reward is at least $V_0-\sqrt{V_0}\,W(H\setminus H_J)$, and the total scaled share is at most $W(H_J)/\sqrt{V_0}\le1$. Multiplying these shares by $B_0$ gives feasible shares in the original problem. The principal's utility is at least
\begin{align*}
&\left(1-\frac{W(H_J)}{\sqrt{V_0}}\right)
\left(V_0-\sqrt{V_0}\,W(H\setminus H_J)\right)\\
&\qquad=V_0-\sqrt{V_0}\,W(H)+W(H_J)W(H\setminus H_J)
\ge\psi_H(V_0).
\end{align*}
There are at most $|H|$ deletions. The threshold comparisons can be made by squaring their nonnegative sides, so the algorithm need not compare sums of square roots. Finally, $\psi_H$ is zero up to $W(H)^2$, and its derivative above that point is $1-W(H)/(2\sqrt v)\in[1/2,1]$. This proves the stated monotonicity and Lipschitz bound.
\end{proof}

For each searched multiplier $\theta$ satisfying $\theta\sum_{i\in L}k_i<1$, the deletion lemma applies with $B_0=1-s_\theta/\theta$ and $V_0=V_\theta$. It guarantees utility at least $\psi_H(V_\theta)\ge\psi_H(\mathcal L(\theta))$. We first bound this expression at the ideal multiplier $\theta^*$, then account for the grid error. By \eqref{eq:alg-root-bounds},
\[
\psi_H(\mathcal L(\theta^*))\ge\left[\mathcal L(\theta^*)-
\sqrt{\left(\sum_{i\in H}k_i\right)\reward(H)\mathcal L(\theta^*)}\right]_+.
\]
The inequality
\[
\sqrt{\left(\sum_{i\in H}k_i\right)\reward(H)\mathcal L(\theta^*)}
\le b\reward(O)\sum_{i\in H}k_i
+\frac{\reward(H)\mathcal L(\theta^*)}{4b\reward(O)}
\]
bounds this additional payment loss. Also, since $\reward(H)\le b\reward(O)$ and $b\le1/2$,
\begin{align*}
&\bigl[b\reward(O)+(1-b)\reward(H)\bigr]
\left(1-\frac{\reward(H)}{4b\reward(O)}\right)-b\reward(O)\\
&\qquad=\reward(H)\left[\frac34-b-
\frac{(1-b)\reward(H)}{4b\reward(O)}\right]\\
&\qquad\ge\reward(H)\left(\frac12-\frac{3b}{4}\right)\ge0.
\end{align*}
Combining these bounds with \eqref{eq:alg-restored-baseline} gives
\begin{align}\label{eq:alg-lifting-fraction}
&\mathcal L(\theta^*)-
\sqrt{\left(\sum_{i\in H}k_i\right)\reward(H)\mathcal L(\theta^*)}\notag\\
&\qquad\ge\left(1-\frac{\reward(H)}{4b\reward(O)}\right)\mathcal L(\theta^*)
-b\reward(O)\sum_{i\in H}k_i\notag\\
&\qquad\ge b\bigl(1-\kappa(O)\bigr)\reward(O).
\end{align}
Thus the small-agent approximation fraction also covers the large agents' payments.

\paragraph{Why a finite grid is enough.}
Choose the largest grid point $\theta\le\theta^*$. Then
\[
\frac{\theta^*}{1+\tau/12}\le\theta\le\theta^*,\qquad
0\le\frac1\theta-\frac1{\theta^*}\le\frac\tau{12},
\]
and $\theta\sum_{i\in L}k_i<1$. The exact gap in the Cauchy--Schwarz bound above is
\begin{align*}
\mathcal L(\theta^*)-\mathcal L(\theta)
&=\left(
\sqrt{\reward(H)\left(\frac1\theta-\sum_{i\in L}k_i\right)}
-\sqrt{b\bigl(\reward(O)-\reward(H)\bigr)\left(1-\frac1\theta\right)}
\right)^2\\
&\le\left(\sqrt{\reward(H)}+
\sqrt{b\bigl(\reward(O)-\reward(H)\bigr)}\right)^2
\left(\frac1\theta-\frac1{\theta^*}\right)\\
&\le\frac{\tau\reward(O)}6.
\end{align*}
For the middle inequality, the squared expression vanishes at $\theta^*$, and
$|\sqrt x-\sqrt y|\le\sqrt{|x-y|}$. The last inequality uses
$b\bigl(\reward(O)-\reward(H)\bigr)+\reward(H)\le\reward(O)$.
The function $\psi_H$ in \Cref{lem:alg-reserved-pruning} is nondecreasing and $1$-Lipschitz, so this grid error loses at most $\tau\reward(O)/6\le\tau\OPT(\reward,\cost)/2$ in actual utility. Let $S_{\rm alg}$ be the team returned by the algorithm. Together with \eqref{eq:alg-lifting-fraction}, the grid bound proves
\begin{equation}\label{eq:alg-final-fraction}
\revenue(S_{\rm alg})\ge
\left(b-\frac\tau2\right)\OPT(\reward,\cost)
=(\gamma_\star-\tau)\OPT(\reward,\cost).
\end{equation}

The algorithm enumerates at most $n^{O(1/\delta)}$ large-agent sets and $O(1/\tau)$ multipliers. Its reward-scale interval has ratio at most $6n$. Since \Cref{lem:alg-diffuse} gives $\eta$ as an inverse polynomial in $1/\tau$, each subroutine and the whole enumeration use $n^{\mathrm{poly}(1/\tau)}$ value queries and arithmetic operations.

\paragraph{Completion of \Cref{thm:computation}.}
For the given $\varepsilon>0$, choose
\[
\tau=\min\left\{\frac1{10},\frac{\varepsilon}{20}\right\}.
\]
This choice is rational for rational $\varepsilon$. Since $\gamma_\star>2/5$, it satisfies $\tau\le\gamma_\star/2$ and $2\tau/\gamma_\star^2\le\varepsilon$. The statewise guarantee \eqref{eq:alg-final-fraction} gives
\[
\frac1{\gamma_\star-\tau}-\frac1{\gamma_\star}
=\frac{\tau}{\gamma_\star(\gamma_\star-\tau)}
\le\frac{2\tau}{\gamma_\star^2}\le\varepsilon,
\]
which proves the claimed approximation when the state is known.

Under uncertainty, reveal the realized state $\sstate$ and run this algorithm on $(\reward_\sstate,\cost)$ with accuracy $3\varepsilon/4$. Its expected utility is at least
\[
\frac{\expect[\sstate\sim\prior]{\OPT(\reward_\sstate,\cost)}}
{1/\gamma_\star+3\varepsilon/4}
=\frac{U_P^F(\prior)}{1/\gamma_\star+3\varepsilon/4}
\ge\frac{\OPT(\prior)}{4/(3\gamma_\star)+\varepsilon},
\]
where the last inequality is \Cref{thm:submodular}. This gives the second approximation factor. The contract rule uses $n^{\mathrm{poly}(1/\varepsilon)}$ value queries for each revealed state and requires no enumeration of the state space.

\end{document}